\documentclass[conference]{IEEEtran}
\usepackage{proj}

\title{Enforcing Application-Layer Policies in eBPF}

\author{
  \IEEEauthorblockN{Laurin Brandner \orcidicon{0009-0005-8251-9117}}
  \IEEEauthorblockA{\text{ETH Zürich} \\
    laurinb@ethz.ch}
  \and
  \IEEEauthorblockN{Ayush Mishra \orcidicon{0000-0001-9723-4513}}
  \IEEEauthorblockA{\text{ETH Zürich} \\
    aymishra@ethz.ch}
  \and
  \IEEEauthorblockN{Sebastiano Miano \orcidicon{0000-0002-1247-9640}}
  \IEEEauthorblockA{\text{Nvidia} \\
    smiano@nvidia.com}
  \linebreakand 
  \IEEEauthorblockN{Aurojit Panda \orcidicon{0000-0001-9664-4377}}
  \IEEEauthorblockA{\text{NYU} \\
    apanda@cs.nyu.edu}
  \and
  \IEEEauthorblockN{Gianni Antichi \orcidicon{0000-0002-6063-4975}}
  \IEEEauthorblockA{\text{Politecnico di Milano} \\
    gianni.antichi@polimi.it}
  \and
  \IEEEauthorblockN{Laurent Vanbever \orcidicon{0000-0003-1455-4381}}
  \IEEEauthorblockA{\text{ETH Zürich} \\
    lvanbever@ethz.ch}
}

\begin{document}

\maketitle

\begin{abstract}
Service meshes have recently emerged as the de-facto standard for deploying microservices. Conceptually, they provide a uniform abstraction for inter-process communication (IPC) between services by implementing common networking mechanisms---such as encryption, routing, and load balancing---and by allowing these mechanisms to be configured and composed through high-level policies. Supporting these policies, however, comes with a significant performance cost, since service meshes interpose proxies (``sidecars'') on the data path between every service.

This paper presents Beeline, a fast path for service meshes which can enforce the vast majority of application-layer policies seen in the wild directly in kernel space. Given high-level policies, Beeline automatically synthesizes an eBPF-based data plane which enforces them in the kernel. Beeline accelerates existing microservices without any code modification, and transparently falls back to existing service proxies (the slow path) for the few unsupported policies.

We fully implemented Beeline, with support for both TLS and HTTP/2. Compared to state-of-the-art service meshes, Beeline reduces the median request latency of realistic applications by up to $6\times$ while sustaining $3\times$ more throughput.
\end{abstract}


\section{Introduction}

Modern data center applications are no longer monolithic: they are assembled from swarms of microservices~\cite{benchara_efficient, jha_study, gan_deathstarbench, kakivaya_servicefabric} or ``pods'', stitched together by service meshes~\cite{saleh_an, li_service}. Service meshes simplify development by abstracting away the networking layer, making applications easier to deploy and manage. This approach has fueled the widespread adoption of platforms such as Istio ~\cite{online_istio} and Linkerd~\cite{online_linkerd}.

A key component of these service meshes is the service proxy, also known as ``{\em sidecar}''. A service proxy acts as the data plane of the service mesh: it processes all the inter-pod traffic and enforces policies defined at the transport (L4) and/or application layer (L7). For example, Istio relies upon Envoy~\cite{online_envoy}---a popular service proxy---to enforce L4 load balancing or L7 request authorization.

\begin{figure}[t]
    \centering
    \input{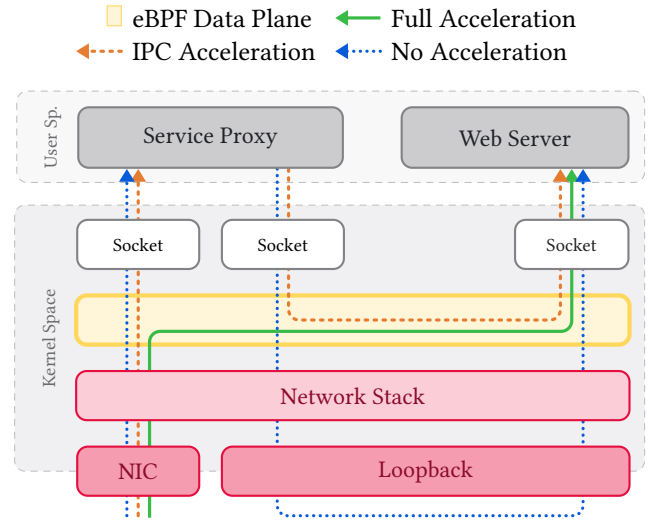}
    \caption{Service proxies typically route inter-pod traffic through the loopback device (blue line). State-of-the-art service meshes optimize IPC by rerouting the traffic using eBPF (orange line). \proj enforces L7 policies in the kernel, eliminating the service proxy from the critical path (green line).}
    \label{fig:intro:datapath}
\end{figure}

While convenient, service proxies can slow down a service mesh significantly: previous studies have shown that they can increase request latency by up to 185\%~\cite{zhu_dissecting}, and increase CPU utilization by 41\%--92\%. This overhead can be traced back to two main sources: (1)~Service proxies execute highly general code and can enforce any conceivable policy. This generality simplifies deployment but carries substantial performance overheads. (2)~Service proxies increase the amount of inter-process communication (IPC), which in turn increases both CPU requirements and processing time. We visualize this extra cost in \Cref{fig:intro:datapath} (see the {\color{no-acc-color} blue} line): each message that the web server receives is routed through the loopback device, traversing the network stack and crossing the user-kernel-boundary {\em three}\ times.

To reduce this overhead, state-of-the-art service meshes such as Cilium~\cite{online_cilium} and Calico~\cite{online_calico} have mainly relied upon two techniques. The first technique aims at minimizing the IPC cost by bypassing the network stack using eBPF~\cite{online_ebpf} (see the {\color{l4fp-color} orange} line in \Cref{fig:intro:datapath}). In practice, though, such bypasses tend to yield limited performance improvements because the bottleneck tends to be, as we show in this paper, the message processing time in the service proxy. Acknowledging this limitation, the second technique (illustrated in {\color{l7fp-color} green}) aims at removing the service proxy entirely from the critical path by offloading L4 policies to the kernel~\cite{online_cilium, online_calico}.

While extremely effective, state-of-the-art service meshes still rely on the service proxy to enforce L7 policies because they deem them too complex to enforce within eBPF. Indeed, despite recent advances to improve L7 support in the kernel, such as strparser~\cite{online_strparser} or KCM~\cite{online_kcm}, processing L7 protocols remains impractical within eBPF's stringent limitations. The reason for this is the self-describing structure of many L7 protocols, e.g., \http. Enforcing policies on such protocols requires more complex parsing logic and more state than protocols with an implicit structure.

Unfortunately, L7 policies also constitute the bulk of the policies in modern deployments. As an illustration, Alibaba reports that the vast majority of their customers (between 80\% to 95\%) use L7 policies in their service mesh~\cite{song_canal}.

But there is an opportunity: While many L7 protocols are complex to parse, the logic needed to enforce L7 policies tends to be simple. This observation enables a split design in which the vast majority of L7 policies are enforced safely in the kernel, using eBPF; while the few more complicated L7 policies are transparently handled in user space by existing service proxies. Realizing this split design is challenging as it requires to overcome eBPF's stringent limitations. In this paper, we present a kernel-based fast path---named {\em \proj}---that can transparently accelerate the most common L7 policies (e.g., HTTP-based). It achieves this while jointly addressing both main sources of overhead: generality and IPC cost.

Given an L7 service policy, \proj first automatically synthesizes a highly optimized and policy-specific data plane in eBPF and loads it into the kernel. As we show, \proj-synthesized code is able to process in the kernel the vast majority (89\%) of policies found in 2417 open-source projects. In doing so, \proj completely eliminates all IPC overhead that would typically arise, similarly to L4 offloads. In the unlikely case a policy cannot be enforced in the kernel, \proj transparently falls back to the user space service proxy. We stress that existing service proxies do {\em not}\ need to be modified and that \proj works with \emph{any} service proxy.

We implemented \proj and demonstrate its practicality by speeding up state-of-the-art service proxies such as Envoy, and show that \proj significantly improves performance. Under realistic conditions, \proj reduces the median request latency of state-of-the-art service meshes by up to $6\times$, and the $99^{th}$-percentile by $4\times$, while serving up to $3\times$ more traffic. To sum up, we make the following contributions:

\begin{enumerate}
    \item We make the case for processing L7 policies in kernel space and demonstrate that eBPF is mature enough to implement the most commonly used L7 policies observed in the wild.
    \item We introduce \proj, an eBPF-based fast path for any service proxy that processes L7 policies in user space. It overcomes the drawbacks of state-of-the-art solutions by synthesizing an efficient and L7-capable data plane.
    \item We implement \proj with support for \hone, \htwo, TLS and six popular policies. We evaluate its performance on synthetic and realistic workloads and share the code to foster reproducibility. \footnote{The source code is available at \repo.}
\end{enumerate}

\section{The Case for an L7 Fast Path}
\label{sec:motivation}
In this section, we make the case for enforcing L7 policies directly in the kernel. We begin with a brief background on how popular service proxies such as Envoy~\cite{online_envoy} operate before quantifying their main sources of overhead. We show that, with or without IPC acceleration, Envoy imposes significant performance overheads. We then discuss what it would take to reduce these overheads and use those insights to motivate \proj's design.

\subsection{Background: Service Proxies}

Services proxies act as the data plane of service meshes and are responsible for handling all the inter-pod traffic. Proxies are typically deployed either \emph{per pod} (e.g., a sidecar in each pod) or \emph{per node} (one proxy handling traffic for multiple applications on the same host). A \emph{service policy} specifies how the proxy should handle the traffic. Typical service policies relate to inter-pod routing, protocol bridging, and security (e.g., authorization). 
As such, they tend to operate either at L4 (e.g., forwarding a message from one port to another) or at L7 (e.g., routing an HTTP request to a particular pod based on method, path, or headers).

\subsection{Identifying the Overhead Sources}
To measure the overhead of services proxies, we deploy the \emph{Social Network} application from DeathStarBench~\cite{gan_deathstarbench} while enforcing a minimal L7 policy that routes traffic based on the HTTP path. We measure the average request latency at 1{,}500\,req/s in four different configurations.\footnote{We choose this rate because, in our setup, it exercises the application without exceeding the maximum throughput of all configurations (see \Cref{sec:eval})}

In the first configuration, \emph{Envoy}, we measure the request latency of the application with a single (per-node) instance of Envoy (the {\color{no-acc-color} blue} line in \Cref{fig:intro:datapath}). In the second configuration, \emph{L4 fast path}, we also use Envoy, but accelerate it with an L4 fast path that reduces IPC cost by short-circuiting traffic at the socket layer (the {\color{l4fp-color} orange} line in \Cref{fig:intro:datapath}). In the third configuration, \emph{\proj}, we enforce the L7 policy in the kernel, removing Envoy from the request path (the {\color{l7fp-color} green} line in \Cref{fig:intro:datapath}). We compare these three configurations against the \emph{lower bound}, which represents the application without Envoy, but its traffic accelerated with the L4 fast path.

\begin{table}[t]
\begin{tabular}{p{\tabcolwidth{0.27}} >{\raggedright\arraybackslash}p{\tabcolwidth{0.73}}}
\hline
\textbf{Component} & \textbf{Description} \\ \hline
Parsing                              & Parsing \http in the service proxy. \\ \hline
\makecell[lt]{Policy \\ Enforcement} & Processing the message and enforcing the policy. \\ \hline
IPC                                  & I/O syscalls like \texttt{send} and \texttt{recv}. \\ \hline
Other                                & Remaining portion of the request latency, that consists mostly of the application's processing time. \\ \hline
\end{tabular}
\caption{Generality makes parsing and policy enforcement inefficient. IPC costs arise because the policy is enforced outside of the pod's process.}
\label{tab:motivation:dissect}
\end{table}

We decompose the per-request latency into the three major sources of overhead identified by MeshInsight~\cite{zhu_dissecting}: protocol parsing, policy enforcement, and IPC.\footnote{To reduce attribution error, we instrument the proxy with in-process atomic counters rather than relying on eBPF; the additional measurement overhead is negligible relative to the reported magnitudes.} We label the remaining time that consists mostly of the application's processing time as \emph{Other} (c.f. \Cref{tab:motivation:dissect}). \Cref{fig:motivation:dissect} shows the result.

Envoy adds 3.4\,ms per request, which is 47\% of the overall 7.2\,ms average request latency at this load (lower bound without Envoy: 3.7\,ms, on our testbed described in \Cref{sec:eval}). Adding the L4 fast path reduces per-request latency by 17\%. Across these configurations, we see that Envoy's overhead is dominated by protocol parsing, followed by IPC. 

Protocol parsing accounts for 1.8\,ms (54\%) of Envoy's total overhead. Moreover, even with a minimal policy, enforcement makes up another 7\%. On the other hand, by employing a specialized data plane in kernel space, \proj reduces the total overhead to 0.12\,ms—a 96\% reduction.

Similarly, IPC accounts for 0.5\,ms (15\%) of Envoy's overhead. The L4 fast path reduces this cost by bypassing the network stack—also shrinking the \emph{Other} remainder—for an overall reduction of 17\%. Again, by enforcing the policy in the kernel, \proj eliminates any proxy-induced IPC.


\subsection{Addressing the Major Overheads}
The experiment above illustrates that the performance of service proxies, even for simple L7 policies, is shaped by multiple factors. Protocol parsing dominates overall cost, while IPC overheads make a secondary but measurable contribution. Maximizing performance, therefore, requires jointly optimizing both aspects.

\begin{figure}[t]
    \vspace{-1em}%
    \begin{subfigure}[t]{0.5\linewidth}
        \tikzsetnextfilename{legend-dissect-figure}

\begin{tikzpicture}
    \begin{axis}[
    ybar stacked,
    legend columns = 4,
    legend cell align={left},
    legend style={at={(0,0)},draw=none,anchor=south west, /tikz/every even column/.append style={column sep=0.25cm}},
    hide axis,
    xmax=1, ymax=1,
    ]
        \addplot[fill=uchu-pink-1, draw=uchu-pink-6] coordinates { (0, 0) };
        \addplot[fill=uchu-red-1, draw=uchu-red-6] coordinates { (0, 0) };
        \addplot[fill=uchu-purple-1, draw=uchu-purple-6] coordinates { (0, 0) };
        \addplot[pattern=north west lines, draw=uchu-gray-6, pattern color=uchu-gray-6] coordinates { (0, 0) };

        \legend{Parsing, Policy Enforcement, IPC, Other}
    \end{axis}

\end{tikzpicture}
    \end{subfigure}
    \tikzsetnextfilename{dissect-figure}



\begin{tikzpicture}[
    annot/.style={fill=white, text width=20pt, anchor=east, execute at begin node=\setlength{\baselineskip}{0.75em}},
]
\begin{axis}[
ybar stacked,
ymin=0,
ymax=8,
axis lines=left,
bar width=20pt,
ylabel={Latency [ms]},
xticklabels={Envoy, {L4 Fast\\Path}, {\proj}},
xticklabel style={align=center},
xtick=data,
every extra y tick/.style={
        yticklabel style={
            anchor=west,
            fill=white,
            inner sep=0,
            xshift=2.25cm,
        },
},
extra y tick label=\empty,
extra y ticks={2, 3.7277403390787702, 6, 8},
extra y tick style={
    grid=major,
    grid style=dashed,
},
enlarge x limits={abs=1.5cm},
height=4cm,
width=\linewidth]

\addplot[ybar, preaction={fill, white}, pattern=north west lines, draw=uchu-gray-6, pattern color=uchu-gray-6] plot coordinates {(1, 4.536425959321376) (2, 3.790062489828131) (3, 3.7277403390787702) };
\addplot[ybar, draw=uchu-pink-6, fill=uchu-pink-1] plot coordinates {(1, 1.8504876260384118) (2, 1.758893532338398) (3, 0.07775232346050992) };
\addplot[ybar, draw=uchu-red-6, fill=uchu-red-1] plot coordinates {(1, 0.24910943036829625) (2, 0.18706456790234702) (3, 0.042408957338302175) };
\addplot[ybar, draw=uchu-purple-6, fill=uchu-purple-1] plot coordinates {(1, 0.5142542417801036) (2, 0.1586799739108811) (3, 0) };

\node[anchor=south] (l4fp label) at (2, 5.894700564) {\scriptsize - 17\%};
\node[anchor=south] (beeline label) at (axis cs: 3, 3.8479016199) {\scriptsize - 46\%};

\node[annot, align=center] (annot) at (3.7, 3.7) {\scriptsize lower\\bound};
\path[name path=axis, draw=lightgray] (axis cs:0,3.7277403390787702) -- (axis cs:4,3.7277403390787702);

\end{axis}
\end{tikzpicture}
    \caption{Protocol parsing, policy enforcement, and IPC are the main sources of overhead that dictate Envoy's performance. \proj optimizes these inefficiencies simultaneously, resulting in a 46\% lower request latency.}
    \label{fig:motivation:dissect}
\end{figure}
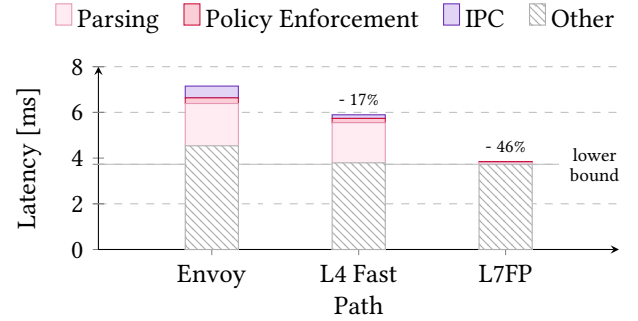

In this work, we propose \proj, a fast path for L7 policies. It specializes the data plane to the policy and thus avoids unnecessary work when parsing the protocol and enforcing the policy. Moreover, it offloads the data plane to kernel space, allowing it to eliminate any IPC overhead that would arise from enforcing the policy outside of the pod's process. Together, these optimizations reduce the Social Network's request latency by 46\% (c.f.~\Cref{fig:motivation:dissect}).

The following paragraphs discuss why \proj's data plane can profit from specialization and what the benefits of offloading that data plane to the kernel space using eBPF are.

\paragraph{Specializing the data plane}
Service proxies like Envoy employ one single, complex data plane that is designed to support every policy on every protocol. This results in complexity that is unnecessary in most cases, which in turn leads to expensive policy enforcement and protocol parsing.

\proj takes the opposite approach: It synthesizes a specialized data plane that is specific to the service policy. Program specialization is a well-known approach that simplifies code by fixing variables to known values~\cite{bhatia_automatic, mcnamee_specialization}. This makes otherwise necessary complexity redundant, allowing \proj to eliminate dead code, use more efficient data structures, and compile policy-specific invariants directly into the binary. Exploiting these optimization opportunities renders the data plane simpler and ultimately, more efficient.


\paragraph{Offloading the data plane} As previously noted, IPC overhead arises because the policy is enforced outside of the pod's process. This results in additional context switches that grow linearly with the call graph size of each request. By offloading the data plane to the kernel, these rising IPC costs can be avoided.
Despite this, the experiment above suggests that a specialized data plane in user space can be useful, too.

Kernel offloading is commonly done with two different techniques: Kernel modules or eBPF~\cite{online_ebpf}. In recent years, eBPF has become increasingly popular as it facilitates kernel offloads significantly. With the Compile Once-Run Everywhere (CO-RE) feature, eBPF achieves portability across different architectures and kernel versions. This simplifies development and lowers maintenance costs. In the context of service proxies, eBPF has another advantage: Its interface makes it easy to dynamically deploy new policies.

However, using eBPF to offload the data plane comes with its own set of limitations. Most importantly, the Linux kernel verifies the safety of an eBPF program by performing an exhaustive execution path traversal~\cite{vishwanathan_verifying, sun_validating} before it is loaded into its address space. This process becomes intractable for programs with complex control flows, hence limiting the complexity that can be offloaded with eBPF. Moreover, the eBPF runtime is event-based, making it impossible for an eBPF-based data plane to send messages on its own accord. This renders the offload of some L7 policies impractical.

\begin{table}[t]
\begin{tabular}{p{\tabcolwidth{0.215}} p{\tabcolwidth{0.785}}}
\hline
\textbf{Category} & \textbf{Description} \\ \hline
\makecell[lt]{Custom}    & Extending Envoy's off-the-shelf functionality with custom modules and external pods. \\ \hline
Protocol                           & Bridging between protocols and protocol-specific data extraction. \\ \hline
Routing                            & Routing, measuring, circuit-breaking, and load balancing traffic. \\ \hline
Security                           & Authenticating, authorizing, encrypting, and signing traffic. \\ \hline
\end{tabular}
\caption{Service proxies are most commonly used to route traffic, but also to enforce security policies or act as bridges between protocols. }
\label{tab:motivation:stats}
\end{table}

A common practice to circumvent these limitations is the modification of the kernel with a kernel module. It is not verified and can execute arbitrarily complex programs. However, as we will show, the most popular L7 policies are simple enough to offload with eBPF. We show this by analyzing the L7 policies used in 2417 open-source projects on GitHub. More specifically, we analyze 4699 distinct Envoy configuration files, and classify all containing L7 policies in one of the categories listed in \Cref{tab:motivation:stats}. We manually inspect each policy to determine if it can be implemented in eBPF without kernel changes. The results are shown in \Cref{fig:motivation:stats}.


Our analysis shows that 89\% of all deployed L7 policies can be implemented without any kernel changes. However, there are some exceptions. 83\% of custom policies extend Envoy's functionality with a custom runtime like WASM, Lua, or Go. Such policies must be transpiled to the eBPF instruction set architecture (ISA) and are unlikely to pass verification without careful consideration. 8\% of the protocol-related policies apply (de-)compression on the HTTP body. 2\% of the routing policies are not event-based. For example, health checking requires the host to send messages to the upstream pods on a regular basis. Finally, 13\% of security policies make use of cryptographic hash functions that are currently not supported by eBPF\footnote{Note that the Linux kernel already implements an extensive crypto API~\cite{online_cryptoapi} and starting from version 6.10, exposes limited functionality to the eBPF runtime. In \Cref{sec:implementation}, we discuss how to extend the existing interface with minimal kernel changes.}. eBPF is a continuously evolving platform with more features added every day. As such, we expect the number of policies that require kernel changes to shrink in the future.

We conclude that eBPF is sufficient to offload the most popular L7 policies to kernel space. In the following section, we discuss how \proj accelerates applications in detail.

\newcommand{\feed}{\texttt{/feed}\xspace}
\newcommand{\admin}{\texttt{/admin}\xspace}

\section{Design}
\label{sec:design}

\proj enforces the vast majority of the L7 policies found in the wild (\Cref{fig:motivation:stats}) directly in the kernel using a streamlined, eBPF-based data plane. For the (few) unsupported policies, \proj automatically and transparently redirects the corresponding requests to existing service proxies, such as Envoy.

In many ways, \proj's design follows a classical SDN split~\cite{mckeown_openflow, pfaff_the} composed of a data plane and a control plane, alongside with an interface between the two. The key difference is the level of abstraction at which \proj functions: \proj's data plane forwards incoming L7 messages onto sockets, rather than packets onto ports.

\begin{figure}[t]
    \vspace{-1em}%
    \begin{subfigure}[t]{0.5\linewidth}
        \tikzsetnextfilename{legend-dissect-figure}

\begin{tikzpicture}
    \begin{axis}[
    ybar stacked,
    legend columns = 2,
    legend cell align={left},
    legend style={at={(0,0)},draw=none,anchor=south west, /tikz/every even column/.append style={column sep=0.25cm}},
    hide axis,
    xmax=1, ymax=1,
    ]
        \addplot[fill=uchu-green-1, draw=uchu-green-6] coordinates { (0, 0) };
        \addplot[fill=uchu-red-1, draw=uchu-red-6] coordinates { (0, 0) };

        \legend{No kernel changes required, Kernel changes required}
    \end{axis}

\end{tikzpicture}
    \end{subfigure}\\
    \begin{subfigure}[b]{\linewidth}
        \tikzsetnextfilename{stats-figure}


\begin{tikzpicture}
\begin{axis}[
ybar stacked,
height=4cm,
ymax=6000,
ymin=0,
width=\linewidth,
ylabel={Occurrences},
axis lines=left,
bar width=20pt,
enlarge x limits={abs=1.2cm},
legend pos=south east,
xticklabels={Custom, Protocol, Routing, Security},
xticklabel style={align=center},
xtick=data,
y tick label style={
    /pgf/number format/fixed,
    /pgf/number format/precision=1,
    /pgf/number format/1000 sep={},
},
yticklabel={\pgfmathparse{\tick/1000}\pgfmathprintnumber{\pgfmathresult}\pgfmathparse{\tick==0 ? "" : "K"}\pgfmathresult},
ymajorgrids=true,
grid style=dashed,
]

\addplot[draw=uchu-green-6, fill=uchu-green-1] coordinates {
(1, 100)
(2, 1199)
(3, 5053)
(4, 1401)
};
\addplot[draw=uchu-red-6, fill=uchu-red-1] coordinates {
(1, 524)
(2, 99)
(3, 82)
(4, 210)
};

\node[anchor=south, text=uchu-red-6] (customization label) at (1, 624) {\scriptsize 83\%};
\node[anchor=south, text=uchu-red-6, fill=white] (protocol label) at (2, 1298) {\scriptsize 8\%};
\node[anchor=south, text=uchu-red-6] (routing label) at (3, 5135) {\scriptsize 2\%};
\node[anchor=south, text=uchu-red-6, fill=white] (security label) at (4, 1611) {\scriptsize 13\%};

\end{axis}
\end{tikzpicture}
    \end{subfigure}
    \caption{89\% of all deployed L7 policies can be implemented in eBPF without kernel changes.}
    \label{fig:motivation:stats}
\end{figure}

\begin{figure}[t]
    \centering
    \input{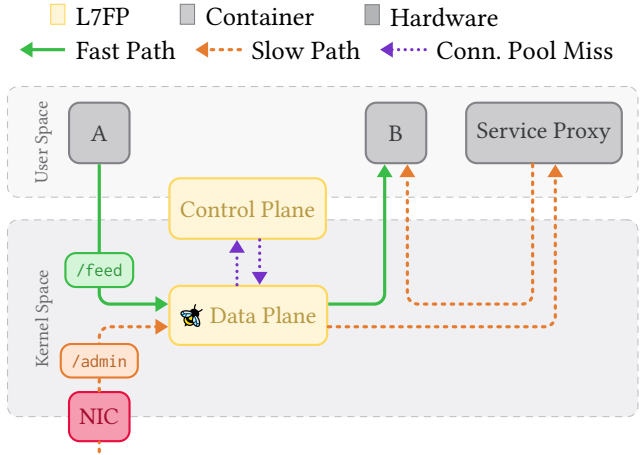}
    \caption{The fast path processes L7 policies in the kernel. The slow path falls back to the service proxy.}
    \label{fig:design:overview}
\end{figure}

In this section we provide an overview of \proj's data and control plane using a running example (\Cref{fig:design:overview}) in which \proj accelerates an application composed of two pods, A and B, and a service proxy. We assume that the application serves two endpoints, \texttt{/feed} and \texttt{/admin}, and that the service mesh is configured with two L7 policies: \bigskip

\begin{noindent}
\begin{tabular}{p{\tabcolwidth{0.25}}p{\tabcolwidth{0.07}} >{\raggedright}p{\tabcolwidth{0.66}}}
\textbf{Match} & \multicolumn{2}{l}{\textbf{Policy}} \tabularnewline
\hline\\
{\tt /feed} & 1. & add \texttt{accept} header \tabularnewline
    & 2. & route to pod B \tabularnewline\tabularnewline
{\tt /admin} & 1. & run a custom authorization script \tabularnewline
    & 2. & route to pod B \tabularnewline
\end{tabular}
\end{noindent}
\bigskip

The first policy adds an \texttt{accept} header to \feed requests before routing them to pod B. The second policy mandates the requests to \admin to be authorized using a custom script running on the service proxy and that \emph{cannot} execute in the kernel, e.g. because it would require an in-kernel interpreter. This is a typical deployment where maintenance endpoints are secured with a company-wide authorization script.

As \Cref{fig:design:overview} shows, \proj eventually ends up enforcing \feed requests entirely in the kernel. Conversely, it ends up routing \admin requests onto an open socket to the service proxy, as these requests cannot be authorized in the kernel.

\subsection{Data Plane}
\label{sec:design:dataplane}

Given a high-level policy, \proj automatically synthesizes an eBPF-based data plane and loads it into the kernel. This data plane adopts the \emph{Parse}-\emph{Match}-\emph{Action} paradigm commonly seen in network functions. It intercepts all (ingress, egress, or local) requests to the application, enforces the corresponding policy, and redirects the traffic to the appropriate socket. \Cref{fig:design:datapath} visualizes the integration into the kernel for the ingress data path. First, at the TCP layer, the kernel manages the TCP state machine and reassembles segments to provide a reliable byte stream. Next, kTLS~\cite{online_ktls} leverages stream parser (strparser)~\cite{online_strparser} to delineate TLS records before decrypting them. The decrypted stream then reaches \proj's data plane at the socket level (\texttt{SK\_SKB} for ingress or egress traffic, \texttt{SK\_MSG} for local traffic). It integrates the \emph{Parse} stage with strparser to delineate application-layer messages. This extracts only the necessary headers to enforce the policies, and stores them into a header vector~$H$. Next, the \emph{Match} stage compares the extracted headers with the policies' match criteria. This yields the policy that should be enforced, along with the forwarding target for the request. Finally, the \emph{Action} stage runs a sequence of actions; simple, generic functions that collectively execute the policy. At the end of each action sequence, the data plane performs a connection pool lookup for a previously established socket to the forwarding target. Conceptually, this connection pool is akin to the forwarding table of an SDN switch. The following paragraphs discuss for each stage first their functionality at runtime, and then how \proj synthesizes them.

\begin{figure*}[t]
\centering
\tikzsetnextfilename{datapath-figure}

\def\innerSep{4pt}
\def\radius{4pt}
\def\componentDistance{0.45cm}
\def\pathOffset{0.15cm}

\begin{tikzpicture} [
    x=6pt,y=8pt,
    uchu/.style 2 args={text=uchu-#1-8, draw=uchu-#1-5, fill=uchu-#1-#2},
    dotted/.style={dash pattern=on 0pt off 2\pgflinewidth, line cap=round},
    dashed/.style={dash pattern=on 2\pgflinewidth off 2\pgflinewidth, line cap=round},
    evenlydashed/.style={dash pattern=on 3\pgflinewidth off 3\pgflinewidth, line cap=round},
    solid line/.style={rounded corners=4pt, uchu-gray-8, very thick},
    arrow/.style={solid line, -{Triangle[scale=0.8]}},
    box/.style={rectangle, rounded corners=4pt, minimum height = 0.8cm, inner sep=4, thick},
    match box/.style={box, uchu={#1}{1}, draw=uchu-#1-6, minimum height=0.3cm, text depth=0},
    action box/.style={box, text=uchu-#1-8, draw=uchu-#1-6, fill=white, minimum height=0.3cm, text depth=0},
    code/.style={box, draw=black, text=black, minimum height=0.3cm, text depth=0, thin},
    label/.style={inner xsep=4pt, align=center, text depth=0pt},
    policy page/.style={thick, fill=white, rounded corners=0.2mm},
    state/.style={thick, circle, uchu={yin}{2}, minimum size=6pt, inner sep=0},
    ks/.style={draw=uchu-gray-5, fill=uchu-gray-1, rounded corners=4pt, dash pattern=on 6\pgflinewidth off 6\pgflinewidth, line cap=round},
    us/.style={draw=uchu-gray-5, fill=uchu-gray-1!50, rounded corners=4pt, dash pattern=on 6\pgflinewidth off 6\pgflinewidth, line cap=round},
    container/.style={box, uchu={yin}{2}, minimum height=0.8cm, minimum width = 0.8cm},
]

    \newcommand{\parse}{
    \begin{tikzpicture}[x=2pt,y=8pt]
    \begin{scope}[name prefix=parse-]
        \node[state] (q1) {};
        \node[state, right=1 of q1, yshift=8pt] (q2) {};
        \node[state, right=1 of q1, yshift=-8pt] (q3) {};
        \node[uchu={yin}{2}, right=1 of q2, yshift=-8pt, minimum height=8pt, minimum width=8pt, rounded corners=4pt, thick] (ctx) {\scriptsize H};

        \draw[solid line, uchu-yin-8] (q1) -- (q2) -- (q3) -- (ctx);
    \end{scope}
    \end{tikzpicture}
    }

    \newcommand{\match}{
    \begin{tikzpicture}[x=6pt,y=8pt]
    \begin{scope}[name prefix=match-]
        \node[match box={green}, anchor=west] (feed match) {\scriptsize\ttfamily /feed};
        \node[match box={orange}, anchor=west, yshift=-16pt] (admin match) {\scriptsize\ttfamily /admin};
    \end{scope}
    \end{tikzpicture}
    }

    \newcommand{\action}{
    \begin{tikzpicture}[x=4pt,y=16pt]
    \begin{scope}[name prefix=action-]
        \node[action box={green}] (get feed) {\scriptsize\ttfamily get};
        \node[action box={green}, right=1 of get feed.east, anchor=west] (write feed) {\scriptsize\ttfamily write};
        \node[action box={green}, right=1 of write feed.east, anchor=west] (get fw feed) {\scriptsize\ttfamily get};
        \node[action box={green}, right=1 of get fw feed.east, anchor=west] (forward feed) {\scriptsize\ttfamily forward};

        \node[action box={orange}, below=1 of get feed.west, anchor=west] (get admin) {\scriptsize\ttfamily get};
        \node[action box={orange}, right=1 of get admin.east, anchor=west] (forward admin) {\scriptsize\ttfamily forward};
    \end{scope}
    \end{tikzpicture}
    }

    \newcommand{\strparser}{
    \begin{tikzpicture}[x=6pt,y=16pt]
    \begin{scope}[name prefix=rx-]
        \node[uchu={yin}{1}, fill=uchu-gray-1, minimum height=15pt, minimum width=18pt, rounded corners=4pt, thick] (queue) {};
        \fill[uchu-gray-1] (-0.1, -8pt) rectangle (4pt, 8pt);

        \node[left=2pt of queue.east, uchu={yin}{1}, minimum height=12pt, minimum width=8pt, rounded corners=4pt, dashed, thick] (seg1) {};
    \end{scope}
    \end{tikzpicture}
    }

    \node[box, uchu={red}{1}] (tcp) {\small TCP};

    \node[right=\componentDistance of tcp, inner sep=0] (strparser tls) {\strparser};
    \node[label, above=2pt of strparser tls] (strparser tls label) {\scriptsize strparser};

    \node[box, uchu={red}{1}, right=\componentDistance of strparser tls] (tls) {\small kTLS};

    \node[right=\componentDistance of tls, inner sep=0] (strparser dp) {\strparser};
    \node[label, above=2pt of strparser dp] (strparser dp label) {\scriptsize strparser};

    \node[right=\componentDistance of strparser dp,inner sep=\innerSep, minimum height=1.5cm] (parse) {\parse};
    \node[label, above=2pt of parse, text=uchu-yellow-9] (parse label) {\small Parse};

    \node[right=1 of parse, inner sep=\innerSep, minimum height=1.5cm] (match) {\match};
    \node[label, above=2pt of match, text=uchu-yellow-9] (match label) {\small Match};

    \node[right=1 of match, inner sep=\innerSep, minimum height=1.5cm] (action) {\action};
    \node[label, above=2pt of action, text=uchu-yellow-9] (action label) {\small Action};

    \node[box, uchu={red}{1}, right=\componentDistance of action, yshift=-8pt, minimum height=0.3] (socket sp) {\scriptsize Socket};
    \node[box, uchu={red}{1}, right=\componentDistance of action, yshift=8pt, minimum height=0.3] (socket a) {\scriptsize Socket};

    \node[container, right=1.5*\componentDistance of socket a, minimum height=0.3] (service a) {\scriptsize A};
    \node[container, right=1.5*\componentDistance of socket sp, minimum height=0.3] (service proxy) {\scriptsize Service Proxy};

    \scoped[on background layer] {
        \node[ks, fit=(tcp) (socket a) (socket sp) (action label), fit margins={left=0.2cm,right=0.1cm,bottom=0.25cm,top=0.25cm}] (ker) {};
        \node[text=uchu-yin-6, right=0.2cm of ker.north west, anchor=north west, yshift=-0.2cm] {\scriptsize Kernel Space};

        \node[us, fit=(ker.north -| service a.west) (ker.south -| service proxy.east), fit margins={left=0.1cm,right=0.2cm,bottom=0,top=0}] (usr) {};
        \node[text=uchu-yin-6, left=0.2cm of usr.north east, anchor=north east, yshift=-0.2cm] {\scriptsize User Space};

        \draw[arrow, uchu-yin-6] (tcp.west) -- (strparser tls.west);
        \draw[arrow, uchu-yin-6] (strparser tls.east) -- (strparser dp.west);

        \node[uchu={yellow}{1}, rounded corners=4pt, thick, fit=(parse) (match) (action label) (action), inner sep=\innerSep] (dp) {};
        \node[uchu={yellow}{1}, fill=white, rounded corners=4pt, inner sep=0, fit=(parse)] (parse bg) {};
        \node[uchu={yellow}{1}, fill=white, rounded corners=4pt, inner sep=0, fit=(match)] (match bg) {};
        \node[uchu={yellow}{1}, fill=white, rounded corners=4pt, inner sep=0, fit=(action)] (action bg) {};

        \draw[solid line, uchu-yin-6] (strparser dp.east) -- ($(parse.west) + (1, 0)$);
        \draw[arrow, l7fp-color] ($(parse.east) - (1, 0)$) to[out=0, in=180, in looseness=1.5] ($(match.west) + (\innerSep, 8pt)$) -- ($(service a.west)$);
        \draw[arrow, l4fp-color, dashed] ($(parse.east) - (1, 0)$) to[out=0, in=180, in looseness=1.5] ($(match.west) + (\innerSep, -8pt)$) -- ($(service proxy.west)$);
    }

\end{tikzpicture}
\caption{The ingress data path starts with the TCP layer, which assembles packets into a byte stream. kTLS uses strparser to perform message delineation before decrypting TLS records. \proj uses strparser to delineate HTTP messages. It parses the message, matches it against the configured policy, before executing the appropriate actions and forwarding it to the target.}
\label{fig:design:datapath}
\end{figure*}

\paragraph{Parse} This stage extracts policy-relevant information from each message. It consists of multiple, protocol-specific deterministic finite automaton (DFA). The data plane keeps track of the protocol each connection is currently using, and selects the DFA accordingly. It uses this DFA to parse the data in the receive queue and identify individual messages. It is possible that this data does not yet contain an entire message. In that case, the data plane instruct strparser~\cite{online_strparser} to wait for more data. If a message can be identified, it extracts all policy-relevant information from the message. To this end, the data plane feeds the message byte-by-byte to the DFA. The DFA then indicates the relevant byte ranges to the data plane, which it stores into the header vector~$H$. Data segments that are not relevant can be skipped efficiently. This data structure holds the extracted information and other metadata of \emph{one} L7 message.

\proj constructs each DFA on startup for the given policy. It encodes them as a matrix of integers and loads them into the data plane. Additionally, it defines a header vector that can hold all policy-relevant information. The procedures that execute the DFA remain the same across service policies.

In the example above, the data plane in \Cref{fig:design:datapath} only needs to know the HTTP path to enforce either policy. Thus, the header vector consists of a single pointer to the located path, along with metadata like the length of the header block.

\paragraph{Match} This stage selects the appropriate policy to enforce. It consists of a sequence of comparisons between the header vector~$H$ and the policies' match criteria. The service policy determines the order of this sequence, which encodes the priority for each endpoint's policy. \proj synthesizes this stage as a sequence of \texttt{if} clauses.

In the example above, the \emph{Match} stage in \Cref{fig:design:datapath} employs two \texttt{if} clauses that compare the extracted HTTP path in the header vector with ``\feed'', and ``\admin'', respectively.

\begin{table}[t]
    \begin{tabular}{p{\tabcolwidth{0.3}}p{\tabcolwidth{0.7}}}
    \hline
    \textbf{Action} & \textbf{Description} \\ \hline
    \texttt{compare} & Compares two data values. \\\hline
    \texttt{read/write} & Reads/writes a segment of the message. \\\hline
    \texttt{en-/decode} & En-/decodes data with a given scheme. \\\hline
    \texttt{en-/decrypt} & En-/decrypts data with a given scheme. \\\hline
    \texttt{hash} & Hashes data with a given scheme. \\\hline
    \texttt{get/set} & Manage state in a global data structure. This makes it possible to share state between flows, or read it from user space. \\\hline
    \texttt{forward} & Forwards the message to a downstream, upstream, or proxy socket. \\\hline
    \texttt{drop} & Drops the message. \\\hline
    \end{tabular}
\caption{Actions are the building blocks of L7 policies.}
\label{tab:design:actions}
\end{table}

\paragraph{Action} This stage executes a sequence of actions required by the policy. Each action is a simple function, as listed in \Cref{tab:design:actions}. They are generic and do not change across policies. Every action sequence terminates with a \texttt{drop} or \texttt{forward} action. The latter action queries the connection pool for an open connection that it can use to forward the message.

\proj synthesizes this stage by first identifying which policies it can offload to kernel space. It replaces unsupported policies with a route policy to fall back to the service proxy. Next, it generates code for each policy using simple, restricted eBPF templates that define an action control flow with policy-specific parameters as placeholders. During synthesis, \proj replaces these placeholders with the actual data from the policy. It inserts the resulting code into the data plane code, along with the implementation of each action. This yields the final eBPF code, that can be compiled and loaded into the kernel. Because each template passes eBPF verification, the resulting data plane does, with some limitations (c.f. \Cref{sec:discussion}) , too.

In the example above, for the \feed request, the data plane adds the \texttt{accept} header with \texttt{get} and \texttt{write} before routing it with \texttt{get} and \texttt{forward}. Likewise, for the \admin request, it performs \texttt{get} and \texttt{forward} to route it to the service proxy.

\begin{figure}[t]
    \centering
    \tikzsetnextfilename{synthesis}

\def\innerSep{4pt}
\def\radius{4pt}
\def\pathOffset{0.15cm}
\def\chevronSep{1pt}

\begin{tikzpicture} [
    node distance=0.5cm,
    uchu/.style 2 args={text=uchu-#1-8, draw=uchu-#1-5, fill=uchu-#1-#2},
    dotted/.style={dash pattern=on 0pt off 2\pgflinewidth, line cap=round},
    dashed/.style={dash pattern=on 2\pgflinewidth off 2\pgflinewidth, line cap=round},
    evenlydashed/.style={dash pattern=on 3\pgflinewidth off 3\pgflinewidth, line cap=round},
    solid line/.style={rounded corners=4pt, uchu-gray-8, very thick},
    arrow/.style={solid line, -{Triangle[scale=0.8]}},
    box/.style={rectangle, rounded corners=4pt, minimum height = 2.5425cm, inner sep=4, thick},
    picture/.style={x=4pt, y=4pt, minimum height=0, minimum width=0, inner sep=0pt},
    placeholder/.style={box, minimum height = 0, inner sep=1.5, rounded corners=2pt},
    component/.style={box, uchu={yin}{1}, text=black, fill=white},
    label/.style={inner xsep=4pt, align=center, text depth=0pt},
]

\newcommand{\policy}{
\begin{tikzpicture}[picture]
\begin{scope}[name prefix=policy-]
    \node (title) {\scriptsize\tt\textbf{/feed}};
    \node[below=3 of title.west, anchor=west] (add title) {\scriptsize\tt Mutation:};
    \node[below=2 of add title.west, anchor=west] (add label) {\scriptsize\tt add};
    \node[right=1 of add label, placeholder, uchu={blue}{1}] (add) {\scriptsize\tt accept: */*};
    \node[below=3 of add label.west, anchor=west] (route title) {\scriptsize\tt Route:};
    \node[below=2 of route title.west, anchor=west] (route label) {\scriptsize\tt to};
    \node[right=1 of route label, placeholder, uchu={purple}{1}] (route) {\scriptsize\tt 172.18.0.3};
\end{scope}
\end{tikzpicture}
}

\newcommand{\addTemplate}{
\begin{tikzpicture}[picture]
\begin{scope}[name prefix=template-]
    \node (idx def) {\scriptsize\tt u32 idx};
    \node[below=2 of idx def.west, anchor=west] (idx call) {\scriptsize\tt = get(hdr\_vec, "len");};

    \node[below=2 of idx call.west, anchor=west] (hdr def) {\scriptsize\tt char *hdr = "};
    \node[right=0 of hdr def, anchor=west, placeholder, uchu={blue}{1}, fill=white] (hdr ph) {\scriptsize\tt HDR};
    \node[right=0 of hdr ph, anchor=west] (hdr end) {\scriptsize\tt \textbackslash r\textbackslash n";};

    \node[below=2 of hdr def.west, anchor=west] (write call) {\scriptsize\tt write(idx, hdr);};
\end{scope}
\end{tikzpicture}
}

\newcommand{\routeTemplate}{
\begin{tikzpicture}[picture]
\begin{scope}[name prefix=template-]
    \node (get def) {\scriptsize\tt struct sock *dest};
    \node[below=2 of get def.west, anchor=west] (get call) {\scriptsize\tt = get(conn\_pool, };
    \node[right=0 of get call, anchor=west, placeholder, uchu={purple}{1}, fill=white] (get ph) {\scriptsize\tt IP};
    \node[right=0 of get ph, anchor=west] (get end) {\scriptsize\tt);};

    \node[below=2 of get call.west, anchor=west] (forward call) {\scriptsize\tt forward(dest); };
\end{scope}
\end{tikzpicture}
}

\newcommand{\program}{
\begin{tikzpicture}[picture]
\begin{scope}[name prefix=program-]
    \node (get) {\scriptsize\tt get \{...\}};
    \node[below=2 of get.west, anchor=west] (write) {\scriptsize\tt write \{...\}};
    \node[below=2 of write.west, anchor=west] (forward) {\scriptsize\tt forward \{...\}};
    \node[below=2 of forward.west, anchor=north west] (feed start) {\scriptsize\tt feed \{};
    \node[below=2 of feed start.west, anchor=west, xshift=4pt, placeholder, uchu={blue}{1}] (hdr mut) {\scriptsize\tt mutation};
    \node[below=2.5 of hdr mut.west, anchor=west, placeholder, uchu={purple}{1}] (route) {\scriptsize\tt route};
    \node[below=2 of route.west, anchor=west, xshift=-4pt] (feed end) {\scriptsize\tt \}};
\end{scope}
\end{tikzpicture}
}

\node[component] (policy) {\policy};
\node[label, above=3pt of policy] (policy label) {Service Policy};

\node[component, uchu={blue}{1}, right=5pt of policy.north east, minimum height=0, anchor=north west] (add template) {\addTemplate};
\node[component, uchu={purple}{1}, below=5pt of add template, minimum height=0, minimum width=3.13cm] (route template) {\routeTemplate};
\node[label, above=3pt of add template] (template label) {Template};

\node[component, right=5pt of add template.north east, anchor=north west] (program) {\program};
\node[label, above=3pt of program] (program label) {Data Plane};

\draw[arrow, dashed, uchu-blue-5] ($(policy.east) - (0.15cm, -0.025cm)$) -- ($(add template.west) - (0, 0.3cm)$);
\draw[arrow, dashed, uchu-blue-5] ($(add template.east) - (0, 0.3cm)$) -- ($(program.west) + (0.295cm, -0.3cm)$);

\draw[arrow, dashed, uchu-purple-5] ($(policy.east) - (0.45cm, 0.675cm)$) -- (route template.west);
\draw[arrow, dashed, uchu-purple-5] (route template.east) -- ($(program.west) + (0.295cm, -0.65cm)$);

\end{tikzpicture}
    \caption{\proj synthesizes the \emph{Action} stage with predefined eBPF templates.}
    \label{fig:design:synthesis}
\end{figure}
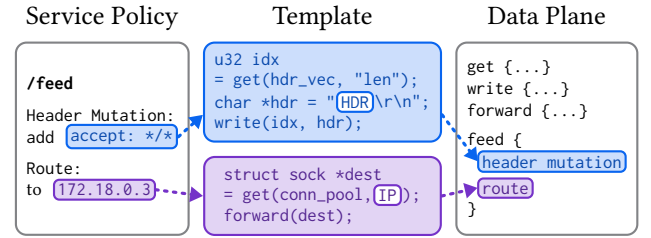

\Cref{fig:design:synthesis} visualizes the synthesis of the \feed policy. It shows that the service policy requires \proj to add an \texttt{accept} header before routing the request to \texttt{172.18.0.3}, the IP address of pod B. \proj uses the header mutation policy to append the new header. This policy first calls the \texttt{get} action to retrieve the insert location of the new header, and then calls \texttt{write} to insert the string into the message. Similarly, it uses the route policy to forward the message to pod B. This policy also first calls \texttt{get} to query the connection pool, and then calls \texttt{forward} to perform the redirection.
\proj replaces the \texttt{accept} header with the \texttt{HDR} placeholder in the header mutation policy, and \texttt{172.18.0.3} with the \texttt{IP} placeholder in the route policy. This results in code that is almost ready to compile. In a final step, \proj concats the specialized templates into a single function, and inserts that function into the data plane code. The data plane code contains the implementation of the actions and the execution engine for the parsing stage.

The synthesis of the \admin policy looks very similar. Because \proj cannot execute custom scripts in eBPF, it replaces the policy with a route policy, into which it inserts the IP address of the service proxy.

Note that policy templates may be L7 protocol specific. In the case of \htwo, it is also necessary for \proj to be able to upgrade connections from \hone, i.e. switch between policy implementations. To account for this, \proj generates the header mutation code with two templates, keeps state at runtime on the currently used protocol, and calls the appropriate action sequence accordingly. Note that header mutation template shown in \Cref{fig:design:synthesis} is \hone specific. This can be seen by the plain text \texttt{hdr} variable that terminates with ``\texttt{\textbackslash r\textbackslash n}''.

\begin{table*}[!htp]
    \begin{tabular}{p{\tabcolwidth{0.1}}p{\tabcolwidth{0.21}}p{\tabcolwidth{0.69}}}
    \hline
    \textbf{Policy} & \textbf{Actions} & \textbf{Description} \\ \hline
    Route & \makecell[lt]{\tt get, forward} & Redirects the message based on the HTTP headers. \\\hline
    \makecell[lt]{Load\\Balancer} & \makecell[lt]{\tt read, hash, get,\\\tt forward} & Load balances a message using the Ketama~\cite{online_ketama} scheme. It hashes a message property, e.g. an HTTP header, with xxHash~\cite{online_xxhash} and uses it to index a list of upstream pods. \\\hline
    JWT & \makecell[lt]{\tt read, decode, hash,\\\tt get, compare, drop} & Authenticates JSON Web Tokens (JWT) with the HS256 authentication scheme and authorizes the request based on the \texttt{issuer} and \texttt{audience} claim. \\\hline
    RBAC & \makecell[lt]{\tt read, get, compare,\\\tt drop} & Enforces network-wide Role Based Access Control (RBAC) policies by white-listing port ranges, source IPs, HTTP paths, etc., for a specific endpoint. \\\hline
    Telemetry & {\tt read, get, set} & Records statistics of the requests and responses. \\\hline
    Mutation & \makecell[lt]{\tt get, write} & Adds, removes, or modifies existing HTTP headers. \\\hline
    \end{tabular}
\caption{\proj supports six popular policies, implemented for \hone and \htwo.}
\label{tab:impl:supported-policies}
\end{table*}

\subsection{Control Plane}
\label{sec:design:controlplane}

eBPF cannot establish new connections, such that \proj relies on a control plane running in user space to maintain the connection pool. It is empty on startup and lazily replenished during runtime. When a connection pool miss occurs, the \proj data plane automatically forwards the request to the control plane. The control plane in turn establishes a new connection to the forwarding target, inserts it into the connection pool, and forwards the request. These connections remain open and are reused until the peer closes them. In this regard, the control plane manages connections to the service proxy like for any other forwarding target.

The connection pool references connections via sockets. To bypass the network stack for local traffic, \proj also requires the sockets of the forwarding target's process. To this end, the control plane uses the \texttt{SOCK\_OPS} hook point.

\section{Implementation}
\label{sec:implementation}

This section describes the implementation of \proj. We implement the control plane in approximately 4K lines of Rust code, the data plane in 2K lines of C code. In the following paragraphs, we outline the details of the \emph{Parse}-\emph{Match}-\emph{Action} architecture, and some challenges of implementing it in eBPF.

\subsection{The Parse Stage}

We implement the \emph{Parse} stage for \hone and \htwo. \proj is extensible and new protocols can be added by providing it with protocol-specific, glob-like patterns. The DFA construction remains fixed.

\paragraph{Parsing TLS-encrypted traffic} A common use case of service proxies is TLS. Service meshes rely on the proxy to authenticate the application to the client, or authenticate two communicating pods with mutual TLS (mTLS).

\proj terminates TLS and parses encrypted traffic directly in the kernel. We highlight two key implementation details required to achieve this. First, \proj offloads TLS to the kernel with kTLS~\cite{online_ktls}. To this end, the control plane performs the TLS handshake upon connection establishment, before passing the cryptographic connection state to the kernel. Second, \proj attaches the data plane to the \texttt{SK\_MSG} or \texttt{SK\_SKB} hook on each socket. It uses the former to process and redirect local traffic before the kernel encrypts it. It uses the latter to process ingress traffic after the kernel decrypts it, or egress traffic before the kernel encrypts it, respectively.


\subsection{The Match Stage}

\proj's data plane matches HTTP headers with basic string comparison functions. Future versions of \proj can extend this with non-backtracking regex patterns, for example.

\subsection{The Action Stage}

Guided by our policy survey (see \Cref{fig:motivation:stats}), GitHub repositories~\cite{online_cilium_github, online_istio_github} and previous work~\cite{saxena_copper, song_canal}, we implement six policies (c.f.~\Cref{tab:impl:supported-policies}): two routing policies, two security policies, telemetry, and one traffic mutation policy. \proj is extensible, and new policies can be implemented by providing \proj with new policy templates. These templates consist of policy parameter placeholders, \proj actions, and \texttt{bpf\_helper} function calls and tend to be short: The policies show in \Cref{tab:impl:supported-policies} are all implemented in <70 LoC.

\paragraph{Encoding and Hashing} Actions like \texttt{hash}, \texttt{en-/decode}, and \texttt{en-/decrypt} are inherently impractical to implement directly in eBPF. While Linux kernel version 6.10 provides \emph{kfuncs}~\cite{online_kfunc} to en- and decrypt data, similar functions to hash or encode arbitrary data are missing. However, popular policies like ring load balancers rely on non-cryptographic hashes functions to distribute traffic evenly across servers, and JSON Web Token (JWT)~\cite{online_jwt} depend on \texttt{base64url}.

Fortunately, the kernel already implements many commonly used hashing and encoding schemes~\cite{online_cryptoapi, online_xxhash}, the eBPF runtime just cannot access them. \proj employs a minimal kernel module (172 lines of C code) which exposes this functionality to the eBPF runtime.

\paragraph{Routing \htwo} The \htwo protocol compresses headers~\cite{online_hpacks} using a Huffman encoding and a cache that allows the sender to transmit references to previously used headers. While the former is stateless, the latter renders the header compression stateful. This poses a problem, as \proj cannot forward header references to a different receiver without dereferencing them first. To avoid this, \proj requires all pods to disable header caching. This is a common feature of \htwo libraries~\cite{online_http2_rust, online_http2_go}. We will show in \Cref{sec:eval} that despite this, \proj accelerates \htwo significantly. Note that Huffman-encoded headers are fully supported and can remain enabled.

Moreover, \proj routes \htwo stream-wise: The header frame at the start of the stream dictates the forwarding target of the following frames. However, not all frame types have a clear destination. For example, a push promise notifies the peer of a stream prior to sending the header frame. Likewise, the flow control frame can be connection-wide. If the forwarding target is unclear, the data plane forwards the frames to the control plane. Future versions of \proj cache push promises and split up flow control frames.

\paragraph{Managing Connections} \proj manages open connections with the connection pool. It uses an eBPF hash map to store all open connections to a host, e.g., a downstream client or an upstream pod. Moreover, it keeps per-connection state to route responses back to the correct recipient. This design gives the data plane the flexibility to enforce policies like dynamic load balancing, or multiplexing long-lived \htwo connections. The connection managment logic uses the \texttt{SOCK\_OPS} hook to monitor sockets from other processes.

\section{Evaluation}
\label{sec:eval}

In this section, we assess the performance of \proj along three dimensions. First, in \Cref{sec:eval:fp}, we analyze the performance benefits that \proj provides. We show that for a realistic workload, \proj can improve the median request latency of state-of-the-art service meshes by up to $6\times$ while sustaining up to $3\times$ more traffic. Second, in \Cref{sec:eval:sp}, we evaluate \proj's overhead in the worst case, where all traffic must be processed by the service proxy and show that it still outperforms Envoy for requests with headers smaller than 6.8\,kB. For requests that are larger than this, the parsing overhead outgrows the IPC acceleration. Finally, in \Cref{sec:eval:cp}, we examine how \proj scales with increasingly complex policies. We find that while \proj's performance degrades faster than Envoy's, its throughput remains at least 39\% higher, even when Envoy is accelerated with an L4 fast path.

\subsection{Methodology}

\paragraph{Applications} We evaluate \proj with three realistic and one synthetic workload. We use Docker Compose to spawn and configure each application.

\begin{description}
    \item[Social Network:] The Social Network is a realistic application from DeathStarBench~\cite{gan_deathstarbench} that uses Thrift RPC~\cite{slee_thrift}, which we configure to use \hone.
    \item[Media Service:] The Media Service is realistic application from DeathStarBench~\cite{gan_deathstarbench} that uses Thrift RPC, which we configure to use \hone.
    \item[Hotel Reservation:] The Hotel Reservation is realistic application from DeathStarBench~\cite{gan_deathstarbench} that uses gRPC~\cite{online_grpc} over \htwo. We disable header caching when benchmarking \proj, but leave it otherwise enabled. We tested both configurations and found that in our setup, the difference is negligible.
    \item[Echo Service:] The Echo Service is a synthetic application consisting of two pods, one frontend, and a single echo pod. For this workload, we send a 100\,B long \hone request to the frontend which forwards it to the echo pod before responding. This emulates small applications with minor IPC overhead.
\end{description}

Note that DeathStarBench's frontend endpoints only support \hone. We perform the experiments accordingly but emphasize that \proj fully supports \htwo + TLS, too.

\paragraph{Service Proxies} In all experiments, we deploy Envoy~\cite{online_envoy}, a state-of-the-art service proxy typically used with Istio~\cite{online_istio}. However, unlike Istio, we deploy Envoy on a per-node basis, i.e., there is only one service proxy for the entire service mesh. For the evaluation presented in this section, we expect a per-node deployment to always be faster than a per-pod deployment. We did not compare against other approaches~\cite{saokar_servicerouter, chen_remote} that break the service proxy abstraction and enforce policies within the application's process because they are closed-source. \proj also runs on a per-node basis. Its data plane operates from the socket level and bypasses Envoy completely for the supported policies. For an unsupported policy, \proj routes the traffic to Envoy. We compare \proj against two Envoy configurations:

\begin{description}
    \item[Envoy:] This deployment uses Envoy without any acceleration. It routes traffic through the loopback device.
    \item[L4 Fast Path:] This deployment reduces the IPC costs of Envoy with an eBPF program at the socket level. The eBPF program reroutes the traffic so that the network stack is bypassed. This configuration replicates the data path of state-of-the-art service meshes~\cite{online_cilium, online_calico}.
\end{description}

\paragraph{Policies} Based on the policy survey (see \Cref{fig:motivation:stats}), GitHub repositories~\cite{online_cilium_github, online_istio_github} and previous work~\cite{saxena_copper, song_canal}, we
specify a set of policies to benchmark \proj in a diverse environment. It is designed to exercise all components of \proj's data plane and includes application-layer policies like RBAC, JWT, routing, and traffic telemetry.

In \Cref{sec:eval:cp}, we will also evaluate \proj's performance as a function of how complex a policy is. We define a policy as being more complex if it parses and processes a large amount of data, and/or requires more instructions to enforce.


\paragraph{Testbed} We evaluate \proj on two nodes, one that generates the traffic using k6~v1.1.0~\cite{online_k6} and the other that hosts the application. The first is equipped with a 24-Core Intel Xeon CPU E5-2670~v3 (2.3GHz), while the latter has a 20-Core Intel Xeon CPU E5-2670~v2 (2.5GHz). They both have 270GB RAM and run Ubuntu 22.04.5~LTS, Envoy~v1.34.0, and Docker~v28.2.2. The machine that hosts the service is running Linux kernel~v6.16.12. All experiments are performed on bare metal, with TurboBoost and dynamic CPU frequency scaling disabled to reduce measurement variance.

\subsection{How Fast Is the Fast Path?}
\label{sec:eval:fp}

We first assess the \textit{best-case scenario}, where all messages are processed on the fast path (i.e. in the kernel). We deploy the three realistic applications with their own policies: \medskip

\begin{noindent}
\begin{tabular}{p{\tabcolwidth{0.25}}p{\tabcolwidth{0.07}} >{\raggedright}p{\tabcolwidth{0.66}}}
\multicolumn{1}{c}{\textbf{Application}} & \multicolumn{2}{l}{\textbf{Description}} \tabularnewline
\hline\\
\multirow{2}{1.7cm}{Social Network} & 1. & authorize the JWT in the \texttt{Authorization} header \tabularnewline
    & 2. & add the \texttt{x-processed-by} header \tabularnewline
    & 3. & route based on HTTP path \tabularnewline\tabularnewline
\multirow{2}{1.7cm}{Media Service} & 1. & enforce 50 RBAC policies \tabularnewline
    & 2. & add the \texttt{x-processed-by} header \tabularnewline
    & 3. & route based on HTTP path \tabularnewline\tabularnewline
\multirow{2}{1.7cm}{Hotel Reservation} & 1. & record request telemetry \tabularnewline
        & 2. & add the \texttt{x-processed-by} header \tabularnewline
        & 3. & route based on HTTP path \tabularnewline
\end{tabular}
\end{noindent}
\medskip

\begin{figure}[t]
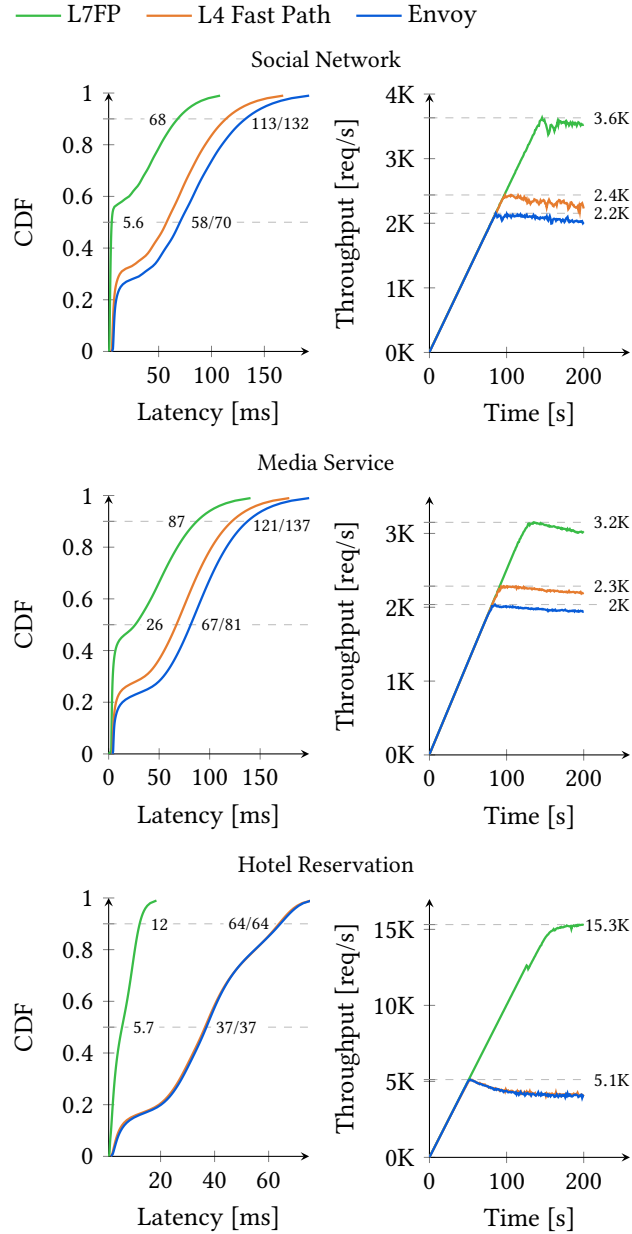

    \vspace{-1em}%
    \begin{subfigure}[t]{0.5\linewidth}
        \tikzsetnextfilename{legend-proxy-figure}

\begin{tikzpicture}
    \begin{axis}[
    legend columns = 4,
    legend style={at={(0,0)},draw=none,anchor=south west, /tikz/every even column/.append style={column sep=0.25cm}},
    hide axis,
    xmax=1, ymax=1,
    ]
        \addplot[beeline] coordinates { (0, 0) };
        \addplot[envoyl4fp] coordinates { (0, 0) };
        \addplot[envoy] coordinates { (0, 0) };

        \legend{\proj, L4 Fast Path, Envoy}
    \end{axis}

\end{tikzpicture}
    \end{subfigure}
    \begin{center}
        \small Social Network
    \end{center}
    \begin{subfigure}[t]{0.5\linewidth}
        \tikzsetnextfilename{fp-sn-cdf-figure}


\begin{tikzpicture}
\begin{axis}[
xlabel={Latency [ms]},
ylabel={CDF},
ymin=0, ymax=1,
axis lines=left,
x tick label style={
    /pgf/number format/fixed,
    /pgf/number format/precision=1,
    /pgf/number format/1000 sep={},
},
legend columns = 4,
legend style={at={(0,1.1)},draw=none,anchor=south west, /tikz/every even column/.append style={column sep=0.25cm}},
scaled x ticks=false,
xtick={0, 50, 100, 150, 200},
xlabel style={anchor=north},
extra y tick label={\empty},
extra y ticks={0.5, 0.9},
extra y tick style={
    grid=major,
    grid style=dashed,
},
grid style=dashed,
height=5cm,
width=\linewidth]

\addplot[beeline] coordinates {
        (2.740147, 0.0)
(3.30361487, 0.01)
(3.38655, 0.02)
(3.44659, 0.03)
(3.496652, 0.04)
(3.54148035, 0.05)
(3.582543, 0.06)
(3.621436, 0.07)
(3.65866196, 0.08)
(3.694436, 0.09)
(3.7291927, 0.1)
(3.763315, 0.11)
(3.79698544, 0.12)
(3.829943, 0.13)
(3.862625, 0.14)
(3.894873, 0.15)
(3.927139, 0.16)
(3.959347, 0.17)
(3.991291, 0.18)
(4.023178, 0.19)
(4.05529, 0.2)
(4.08738827, 0.21)
(4.119848, 0.22)
(4.152386, 0.23)
(4.185226, 0.24)
(4.218579, 0.25)
(4.252158, 0.26)
(4.286564, 0.27)
(4.321201, 0.28)
(4.356775, 0.29)
(4.3932521, 0.3)
(4.4302, 0.31)
(4.4684, 0.32)
(4.507723, 0.33)
(4.54823, 0.34)
(4.590408449999999, 0.35)
(4.634209, 0.36)
(4.679587, 0.37)
(4.727114060000001, 0.38)
(4.776636, 0.39)
(4.828704, 0.4)
(4.883707, 0.41)
(4.94249754, 0.42)
(5.00515441, 0.43)
(5.072479, 0.44)
(5.145163, 0.45)
(5.224529020000001, 0.46)
(5.31209589, 0.47)
(5.409668, 0.48)
(5.52024463, 0.49)
(5.648398, 0.5)
(5.800054370000001, 0.51)
(5.987275720000001, 0.52)
(6.228492, 0.53)
(6.568461960000001, 0.54)
(7.1113928500000005, 0.55)
(8.261247880000003, 0.56)
(11.53244585999994, 0.57)
(15.25560413999999, 0.58)
(18.356420659999998, 0.59)
(22.06888959999999, 0.6)
(24.152956190000005, 0.61)
(25.938978019999983, 0.62)
(28.001575810000002, 0.63)
(30.163649159999988, 0.64)
(31.8355901, 0.65)
(33.21212514, 0.66)
(34.51578186000003, 0.67)
(35.879827399999996, 0.68)
(37.357845299999994, 0.69)
(38.85070669999997, 0.7)
(40.222389939999985, 0.71)
(41.469523, 0.72)
(42.664254, 0.73)
(43.869844, 0.74)
(45.119223749999996, 0.75)
(46.44249984, 0.76)
(47.80799299, 0.77)
(49.15407834000002, 0.78)
(50.45364, 0.79)
(51.74430900000002, 0.8)
(53.07293723, 0.81)
(54.4707198, 0.82)
(55.94353372999999, 0.83)
(57.468968, 0.84)
(59.01940164999999, 0.85)
(60.607667639999995, 0.86)
(62.26941314, 0.87)
(64.04796748000003, 0.88)
(65.954432, 0.89)
(67.97725990000002, 0.9)
(70.11409437, 0.91)
(72.47563244000001, 0.92)
(75.09968993000001, 0.93)
(78.01989175999998, 0.94)
(81.37985054999996, 0.95)
(85.36006252, 0.96)
(90.29844162999997, 0.97)
(97.03723088, 0.98)
(108.1166419099999, 0.99)
    };
\addplot[envoyl4fp] coordinates {
        (4.979435, 0.0)
(5.81171934, 0.01)
(5.953542339999999, 0.02)
(6.06047051, 0.03)
(6.15469836, 0.04)
(6.2421797, 0.05)
(6.3251360199999995, 0.06)
(6.407043, 0.07)
(6.489215, 0.08)
(6.5729495899999995, 0.09)
(6.6587524, 0.1)
(6.74720574, 0.11)
(6.84012608, 0.12)
(6.938312, 0.13)
(7.04312976, 0.14)
(7.1545471, 0.15)
(7.2743, 0.16)
(7.403504, 0.17)
(7.546250119999999, 0.18)
(7.70381023, 0.19)
(7.881729, 0.2)
(8.08647914, 0.21)
(8.32411048, 0.22)
(8.60297882, 0.23)
(8.939056, 0.24)
(9.34710725, 0.25)
(9.844052399999999, 0.26)
(10.45647536, 0.27)
(11.197737320000002, 0.28)
(12.112899199999989, 0.29)
(13.3418547, 0.3)
(15.23373159, 0.31)
(18.830826439999996, 0.32)
(24.572671310000025, 0.33)
(28.27387554000001, 0.34)
(31.295780499999992, 0.35)
(34.603351319999994, 0.36)
(37.66023505, 0.37)
(39.89900552, 0.38)
(41.67790126, 0.39)
(43.25831339999999, 0.4)
(44.7919687, 0.41)
(46.346503, 0.42)
(47.974547339999994, 0.43)
(49.611614560000014, 0.44)
(51.1975302, 0.45)
(52.65760092, 0.46)
(53.99292397, 0.47)
(55.223449800000004, 0.48)
(56.390097260000005, 0.49)
(57.5294695, 0.5)
(58.65227538, 0.51)
(59.7941952, 0.52)
(60.946312060000004, 0.53)
(62.13135286000002, 0.54)
(63.333895150000004, 0.55)
(64.54277384000001, 0.56)
(65.72880838, 0.57)
(66.8910732, 0.58)
(68.02485547999999, 0.59)
(69.12521199999999, 0.6)
(70.20891374, 0.61)
(71.2781841, 0.62)
(72.3534591, 0.63)
(73.43708092, 0.64)
(74.54821815, 0.65)
(75.68807966, 0.66)
(76.8579689, 0.67)
(78.05292852000002, 0.68)
(79.27161946, 0.69)
(80.5101093, 0.7)
(81.75613184, 0.71)
(83.00463995999999, 0.72)
(84.25476440999999, 0.73)
(85.52780295999999, 0.74)
(86.82845775, 0.75)
(88.16555332, 0.76)
(89.55184554, 0.77)
(90.99104263999999, 0.78)
(92.48504505999999, 0.79)
(94.03934699999999, 0.8)
(95.63457377, 0.81)
(97.26839753999998, 0.82)
(98.94860175999999, 0.83)
(100.70004668, 0.84)
(102.52421124999995, 0.85)
(104.45989783999997, 0.86)
(106.52884280999997, 0.87)
(108.73525880000003, 0.88)
(111.09169368000003, 0.89)
(113.60993290000005, 0.9)
(116.34373999000002, 0.91)
(119.37305868000001, 0.92)
(122.79997978000002, 0.93)
(126.6821009, 0.94)
(131.2091867, 0.95)
(136.62388032, 0.96)
(143.39249440000003, 0.97)
(152.65261244, 0.98)
(167.80561648000003, 0.99)
    };
\addplot[envoy] coordinates {
        (5.908758, 0.0)
(6.787855909999999, 0.01)
(6.94939894, 0.02)
(7.07252791, 0.03)
(7.182484, 0.04)
(7.2838527, 0.05)
(7.3814854599999995, 0.06)
(7.478880790000001, 0.07)
(7.57855752, 0.08)
(7.680995, 0.09)
(7.7870697, 0.1)
(7.900436, 0.11)
(8.02284, 0.12)
(8.157297, 0.13)
(8.30736774, 0.14)
(8.47564555, 0.15)
(8.66878252, 0.16)
(8.89636398, 0.17)
(9.16933, 0.18)
(9.50133329, 0.19)
(9.9023356, 0.2)
(10.394948849999999, 0.21)
(11.006623339999999, 0.22)
(11.77361817, 0.23)
(12.770541839999998, 0.24)
(14.13371575, 0.25)
(16.192371380000004, 0.26)
(19.478407939999997, 0.27)
(25.026096040000002, 0.28)
(31.15126247, 0.29)
(34.6734425, 0.3)
(37.794238729999975, 0.31)
(41.210920040000005, 0.32)
(44.47798656000001, 0.33)
(46.909781020000025, 0.34)
(48.80619959999999, 0.35)
(50.4641328, 0.36)
(51.99273668000001, 0.37)
(53.50057244, 0.38)
(55.032655090000006, 0.39)
(56.620993600000006, 0.4)
(58.26769769999999, 0.41)
(59.90073781999999, 0.42)
(61.47234072, 0.43)
(62.9290008, 0.44)
(64.26122585, 0.45)
(65.51867696000001, 0.46)
(66.70920531, 0.47)
(67.87436448, 0.48)
(69.01366025000002, 0.49)
(70.1585055, 0.5)
(71.32456794, 0.51)
(72.52367508, 0.52)
(73.74498591999999, 0.53)
(74.99649266, 0.54)
(76.27262545000002, 0.55)
(77.54421531999999, 0.56)
(78.8004029, 0.57)
(80.02160278, 0.58)
(81.21254807, 0.59)
(82.3784344, 0.6)
(83.53215355, 0.61)
(84.67752556, 0.62)
(85.83687543, 0.63)
(87.01386448, 0.64)
(88.21917959999999, 0.65)
(89.46913512, 0.66)
(90.75539481, 0.67)
(92.08400576000001, 0.68)
(93.44524483, 0.69)
(94.83323139999999, 0.7)
(96.23005823000001, 0.71)
(97.637122, 0.72)
(99.050016, 0.73)
(100.46960068, 0.74)
(101.91322075, 0.75)
(103.3846392, 0.76)
(104.90895335, 0.77)
(106.4974773, 0.78)
(108.17175852, 0.79)
(109.91267880000001, 0.8)
(111.71017157, 0.81)
(113.55973863999999, 0.82)
(115.46461251, 0.83)
(117.43536951999998, 0.84)
(119.48182905, 0.85)
(121.65003320000001, 0.86)
(123.97707416000009, 0.87)
(126.4789378, 0.88)
(129.17945291, 0.89)
(132.08451740000004, 0.9)
(135.20054857999997, 0.91)
(138.62988292, 0.92)
(142.48013993000004, 0.93)
(146.87340962000002, 0.94)
(151.9216600000001, 0.95)
(157.90968632, 0.96)
(165.48403744, 0.97)
(175.70880512000002, 0.98)
(192.3405296599998, 0.99)
    };

    \coordinate (p50 l7fp) at (5.648398, 0.5);
    \coordinate (p50 no-acc) at (70.1585055, 0.5);

    \coordinate (p90 l7fp) at (67.97725990000002, 0.9);
    \coordinate (p90 no-acc) at (132.08451740000004, 0.9);

\end{axis}

\node[annot y] (p50 l7fp annot) at ($(p50 l7fp) + (2pt, 0)$) {\scriptsize 5.6};
\node[annot y] (p50 no-acc annot) at ($(p50 no-acc) + (2pt, 0)$) {\scriptsize 58/70};

\node[annot y, anchor=east] (p90 l7fp annot) at ($(p90 l7fp) - (2pt, 0)$) {\scriptsize 68};
\node[annot y, anchor=north west] (p90 no-acc annot) at ($(p90 no-acc) + (0, 1pt)$) {\scriptsize 113/132};

\end{tikzpicture}
    \end{subfigure}%
    \begin{subfigure}[t]{0.5\linewidth}
        \input{fig/eval/fp-sn-rate}
    \end{subfigure}
    \begin{center}
        \small Media Service
    \end{center}
    \begin{subfigure}[t]{0.5\linewidth}
        \tikzsetnextfilename{fp-ms-cdf-figure}


\begin{tikzpicture}
\begin{axis}[
xlabel={Latency [ms]},
ylabel={CDF},
ymin=0, ymax=1,
axis lines=left,
x tick label style={
    /pgf/number format/fixed,
    /pgf/number format/precision=1,
    /pgf/number format/1000 sep={},
},
legend columns = 4,
legend style={at={(0,1.1)},draw=none,anchor=south west, /tikz/every even column/.append style={column sep=0.25cm}},
scaled x ticks=false,
xtick={0, 50, 100, 150},
xlabel style={anchor=north},
extra y tick label={\empty},
extra y ticks={0.5, 0.9},
extra y tick style={
    grid=major,
    grid style=dashed,
},
grid style=dashed,
height=5cm,
width=\linewidth]

\addplot[beeline] coordinates {
        (0.0, 0.0)
(2.55136961, 0.01)
(2.60406, 0.02)
(2.6441558300000003, 0.03)
(2.67953544, 0.04)
(2.712938, 0.05)
(2.745381, 0.06)
(2.777668, 0.07)
(2.810382, 0.08)
(2.843801, 0.09)
(2.87821, 0.1)
(2.913618, 0.11)
(2.95038632, 0.12)
(2.98858893, 0.13)
(3.02809, 0.14)
(3.069653, 0.15)
(3.11253776, 0.16)
(3.157385, 0.17)
(3.20437, 0.18)
(3.25352059, 0.19)
(3.304911, 0.2)
(3.359472, 0.21)
(3.41648, 0.22)
(3.47767703, 0.23)
(3.54287664, 0.24)
(3.613187, 0.25)
(3.690456, 0.26)
(3.77416447, 0.27)
(3.867096, 0.28)
(3.969801, 0.29)
(4.08507, 0.3)
(4.21420564, 0.31)
(4.359041, 0.32)
(4.52163, 0.33)
(4.70689174, 0.34)
(4.917469, 0.35)
(5.159974, 0.36)
(5.441143140000001, 0.37)
(5.774046, 0.38)
(6.17225379, 0.39)
(6.656958, 0.4)
(7.268538, 0.41)
(8.065359440000002, 0.42)
(9.140028679999991, 0.43)
(10.623556879999999, 0.44)
(12.71007945, 0.45)
(15.494282700000023, 0.46)
(18.728236279999994, 0.47)
(21.737521079999976, 0.48)
(24.238842009999995, 0.49)
(26.346432, 0.5)
(28.215170310000005, 0.51)
(29.924588160000003, 0.52)
(31.55487099, 0.53)
(33.12772266000001, 0.54)
(34.666769950000024, 0.55)
(36.174028480000004, 0.56)
(37.64939277, 0.57)
(39.08534074, 0.58)
(40.487260979999995, 0.59)
(41.854812, 0.6)
(43.18987678, 0.61)
(44.48722594, 0.62)
(45.76463458, 0.63)
(47.01601712, 0.64)
(48.251596, 0.65)
(49.4767509, 0.66)
(50.70426561, 0.67)
(51.92871296000001, 0.68)
(53.166268450000004, 0.69)
(54.41208899999999, 0.7)
(55.671621619999996, 0.71)
(56.9487596, 0.72)
(58.24804231999997, 0.73)
(59.564755680000005, 0.74)
(60.912124000000006, 0.75)
(62.28589139999999, 0.76)
(63.68245264000001, 0.77)
(65.12113214, 0.78)
(66.58704209000001, 0.79)
(68.08778500000001, 0.8)
(69.64298928, 0.81)
(71.25059004, 0.82)
(72.91128193999995, 0.83)
(74.64817148, 0.84)
(76.47934619999998, 0.85)
(78.39543675999998, 0.86)
(80.41532842000001, 0.87)
(82.5743936, 0.88)
(84.85517203, 0.89)
(87.33251710000002, 0.9)
(90.01803094999998, 0.91)
(92.97163660000001, 0.92)
(96.27492625000001, 0.93)
(100.04282036, 0.94)
(104.39273654999997, 0.95)
(109.63546427999992, 0.96)
(116.24307343999999, 0.97)
(125.35884693999988, 0.98)
(140.48885466000007, 0.99)
    };
\addplot[envoyl4fp] coordinates {
        (3.380477, 0.0)
(3.874686, 0.01)
(3.967187, 0.02)
(4.04292386, 0.03)
(4.114648, 0.04)
(4.185997, 0.05)
(4.259371, 0.06)
(4.336325, 0.07)
(4.418639, 0.08)
(4.50691387, 0.09)
(4.6032251, 0.1)
(4.71122773, 0.11)
(4.83526544, 0.12)
(4.98128206, 0.13)
(5.15646634, 0.14)
(5.37276515, 0.15)
(5.6336179600000005, 0.16)
(5.94976085, 0.17)
(6.329697219999998, 0.18)
(6.785171780000001, 0.19)
(7.338908400000001, 0.2)
(8.021230139999997, 0.21)
(8.891815559999998, 0.22)
(10.043124119999996, 0.23)
(11.6203682, 0.24)
(13.84261175, 0.25)
(17.077780060000006, 0.26)
(21.543351460000004, 0.27)
(26.764777480000014, 0.28)
(31.71406014999999, 0.29)
(35.69634249999999, 0.3)
(38.829736329999996, 0.31)
(41.4016756, 0.32)
(43.61447752, 0.33)
(45.582065660000005, 0.34)
(47.3843974, 0.35)
(49.069386279999996, 0.36)
(50.65857166999999, 0.37)
(52.17227324, 0.38)
(53.621485, 0.39)
(55.019772800000005, 0.4)
(56.37165831, 0.41)
(57.68623455999999, 0.42)
(58.9612453, 0.43)
(60.2006662, 0.44)
(61.40983125, 0.45)
(62.595608819999995, 0.46)
(63.75805611999999, 0.47)
(64.89243752, 0.48)
(66.01392752, 0.49)
(67.1151665, 0.5)
(68.20138348, 0.51)
(69.27141112, 0.52)
(70.33474899000002, 0.53)
(71.38988236, 0.54)
(72.43968570000003, 0.55)
(73.49435340000001, 0.56)
(74.54213336, 0.57)
(75.59419296, 0.58)
(76.64147248, 0.59)
(77.6917096, 0.6)
(78.74584482, 0.61)
(79.81446344, 0.62)
(80.88376133000001, 0.63)
(81.96391368, 0.64)
(83.05832485, 0.65)
(84.16311568, 0.66)
(85.28226989000001, 0.67)
(86.41516768, 0.68)
(87.57162946, 0.69)
(88.741007, 0.7)
(89.93203711999999, 0.71)
(91.14870824, 0.72)
(92.38835039, 0.73)
(93.65289591999999, 0.74)
(94.95672975000001, 0.75)
(96.29678120000001, 0.76)
(97.66688974, 0.77)
(99.07597814000002, 0.78)
(100.52908888, 0.79)
(102.0349156, 0.8)
(103.60971549000001, 0.81)
(105.23675641999999, 0.82)
(106.94070330000001, 0.83)
(108.72217327999999, 0.84)
(110.59049394999998, 0.85)
(112.58059986, 0.86)
(114.68350564, 0.87)
(116.94585759999995, 0.88)
(119.36033372, 0.89)
(121.9737598, 0.9)
(124.84478786000007, 0.91)
(128.00716884, 0.92)
(131.56927958, 0.93)
(135.59665815999995, 0.94)
(140.28741725, 0.95)
(145.91904612, 0.96)
(153.03816366000007, 0.97)
(162.82006484, 0.98)
(178.86336319, 0.99)
    };
\addplot[envoy] coordinates {
        (3.84771, 0.0)
(4.42661388, 0.01)
(4.53681576, 0.02)
(4.630005639999999, 0.03)
(4.72015704, 0.04)
(4.8117318000000004, 0.05)
(4.90879, 0.06)
(5.01366916, 0.07)
(5.13196604, 0.08)
(5.26731344, 0.09)
(5.4268234, 0.1)
(5.626634, 0.11)
(5.88730692, 0.12)
(6.235223, 0.13)
(6.686650960000001, 0.14)
(7.2506205999999995, 0.15)
(7.94659164, 0.16)
(8.82464084, 0.17)
(9.96925284, 0.18)
(11.514358199999998, 0.19)
(13.669582600000007, 0.2)
(16.73596568, 0.21)
(21.03920027999999, 0.22)
(26.558842360000014, 0.23)
(32.64484468000001, 0.24)
(38.378317, 0.25)
(43.02410416, 0.26)
(46.6585566, 0.27)
(49.59672228, 0.28)
(52.0986572, 0.29)
(54.2999956, 0.3)
(56.292875, 0.31)
(58.13141144, 0.32)
(59.85368564, 0.33)
(61.46148652, 0.34)
(62.986999, 0.35)
(64.45379275999998, 0.36)
(65.86076356000001, 0.37)
(67.2207712, 0.38)
(68.54307076, 0.39)
(69.8175688, 0.4)
(71.06755516, 0.41)
(72.28174896, 0.42)
(73.47341152, 0.43)
(74.63689732, 0.44)
(75.78262380000001, 0.45)
(76.91023876000001, 0.46)
(78.0229636, 0.47)
(79.11692187999999, 0.48)
(80.20058060000001, 0.49)
(81.270582, 0.5)
(82.33532976, 0.51)
(83.38904532, 0.52)
(84.44732852000001, 0.53)
(85.49646372000001, 0.54)
(86.5428174, 0.55)
(87.59562260000001, 0.56)
(88.64115948, 0.57)
(89.68919676, 0.58)
(90.7483462, 0.59)
(91.8178512, 0.6)
(92.88829116, 0.61)
(93.96279136, 0.62)
(95.04286056000001, 0.63)
(96.13747216, 0.64)
(97.24583679999999, 0.65)
(98.3697678, 0.66)
(99.50698075999999, 0.67)
(100.66103328000001, 0.68)
(101.83368528, 0.69)
(103.0242908, 0.7)
(104.24716028, 0.71)
(105.48726979999999, 0.72)
(106.75927368, 0.73)
(108.05640872000001, 0.74)
(109.389722, 0.75)
(110.76527088, 0.76)
(112.17757852, 0.77)
(113.63665176000002, 0.78)
(115.13184308, 0.79)
(116.6943928, 0.8)
(118.314485, 0.81)
(120.00885611999998, 0.82)
(121.76838456, 0.83)
(123.60874288, 0.84)
(125.56129159999999, 0.85)
(127.62154023999996, 0.86)
(129.82018175999997, 0.87)
(132.17657659999998, 0.88)
(134.70356440000003, 0.89)
(137.47410020000004, 0.9)
(140.49350468, 0.91)
(143.82804492000002, 0.92)
(147.56600075999998, 0.93)
(151.83289611999996, 0.94)
(156.86839439999997, 0.95)
(162.90749768000015, 0.96)
(170.53885147999998, 0.97)
(181.09701444000007, 0.98)
(198.53877047999987, 0.99)
    };

    \coordinate (p50 l7fp) at (26, 0.5);
    \coordinate (p50 no-acc) at (81, 0.5);

    \coordinate (p90 l7fp) at (87, 0.9);
    \coordinate (p90 no-acc) at (137, 0.9);

\end{axis}

\node[annot y] (p50 l7fp annot) at ($(p50 l7fp) + (2pt, 0)$) {\scriptsize 26};
\node[annot y] (p50 no-acc annot) at ($(p50 no-acc) + (2pt, 0)$) {\scriptsize 67/81};

\node[annot y, anchor=east] (p90 l7fp annot) at ($(p90 l7fp) - (2pt, 0)$) {\scriptsize 87};
\node[annot y, anchor=north west] (p90 no-acc annot) at ($(p90 no-acc) + (0, 1pt)$) {\scriptsize 121/137};

\end{tikzpicture}
    \end{subfigure}%
    \begin{subfigure}[t]{0.5\linewidth}
        \input{fig/eval/fp-ms-rate}
    \end{subfigure}
    \begin{center}
        \small Hotel Reservation
    \end{center}
    \begin{subfigure}[t]{0.5\linewidth}
        \tikzsetnextfilename{fp-hr-cdf-figure}


\begin{tikzpicture}
\begin{axis}[
xlabel={Latency [ms]},
ylabel={CDF},
ymin=0, ymax=1,
xmax=75,
axis lines=left,
x tick label style={
    /pgf/number format/fixed,
    /pgf/number format/precision=1,
    /pgf/number format/1000 sep={},
},
legend columns = 4,
legend style={at={(0,1.1)},draw=none,anchor=south west, /tikz/every even column/.append style={column sep=0.25cm}},
dashed/.style={dash pattern=on 8\pgflinewidth off 4\pgflinewidth, line cap=round},
scaled x ticks=false,
xtick={0, 20, 40, 60},
xlabel style={anchor=north},
extra y tick label={\empty},
extra y ticks={0.5, 0.9},
extra y tick style={
    grid=major,
    grid style=dashed,
},
grid style=dashed,
height=5cm,
width=\linewidth]

\addplot[beeline] coordinates {
        (0.780559, 0.0)
(1.063131, 0.01)
(1.125834, 0.02)
(1.178103, 0.03)
(1.228413, 0.04)
(1.2815440500000004, 0.05)
(1.342177, 0.06)
(1.415061, 0.07)
(1.499466, 0.08)
(1.584247, 0.09)
(1.661496, 0.1)
(1.731907, 0.11)
(1.797798, 0.12)
(1.8610337300000004, 0.13)
(1.922701, 0.14)
(1.983568, 0.15)
(2.044288, 0.16)
(2.105592, 0.17)
(2.1674957799999994, 0.18)
(2.230569, 0.19)
(2.294883200000001, 0.2)
(2.360495, 0.21)
(2.427841, 0.22)
(2.497128, 0.23)
(2.568184, 0.24)
(2.641536, 0.25)
(2.716843, 0.26)
(2.794649670000002, 0.27)
(2.8753458800000007, 0.28)
(2.958807, 0.29)
(3.044954, 0.3)
(3.134891, 0.31)
(3.228132, 0.32)
(3.324974, 0.33)
(3.425637, 0.34)
(3.530408, 0.35)
(3.639546, 0.36)
(3.7532567699999997, 0.37)
(3.8719279800000006, 0.38)
(3.995417, 0.39)
(4.124615400000002, 0.4)
(4.259728, 0.41)
(4.40062682, 0.42)
(4.54776, 0.43)
(4.7012882399999985, 0.44)
(4.861399, 0.45)
(5.027338, 0.46)
(5.199319, 0.47)
(5.376391, 0.48)
(5.558327, 0.49)
(5.743548, 0.5)
(5.931116, 0.51)
(6.120237, 0.52)
(6.30924, 0.53)
(6.49695, 0.54)
(6.683081, 0.55)
(6.866877760000001, 0.56)
(7.046814969999998, 0.57)
(7.224058, 0.58)
(7.396538389999996, 0.59)
(7.56583, 0.6)
(7.7310418100000025, 0.61)
(7.89283, 0.62)
(8.050466, 0.63)
(8.20584, 0.64)
(8.357841, 0.65)
(8.507069, 0.66)
(8.654292, 0.67)
(8.799335, 0.68)
(8.942306, 0.69)
(9.083025699999995, 0.7)
(9.223338, 0.71)
(9.362349119999998, 0.72)
(9.500498329999997, 0.73)
(9.638932, 0.74)
(9.777475, 0.75)
(9.916375, 0.76)
(10.055412170000002, 0.77)
(10.19653, 0.78)
(10.339357, 0.79)
(10.484262, 0.8)
(10.632437, 0.81)
(10.78499722, 0.82)
(10.94198343, 0.83)
(11.10354064, 0.84)
(11.271355850000003, 0.85)
(11.446501, 0.86)
(11.630677, 0.87)
(11.82585, 0.88)
(12.03442, 0.89)
(12.259674, 0.9)
(12.50429, 0.91)
(12.775497, 0.92)
(13.080752060000002, 0.93)
(13.432547, 0.94)
(13.853573899999992, 0.95)
(14.380499159999996, 0.96)
(15.085975369999996, 0.97)
(16.173708699999974, 0.98)
(18.47463147999999, 0.99)
    };

\addplot[envoy] coordinates {
        (1.687603, 0.0)
(2.33596798, 0.01)
(2.6280179599999998, 0.02)
(2.892008, 0.03)
(3.15823892, 0.04)
(3.4319999, 0.05)
(3.720356, 0.06)
(4.03641686, 0.07)
(4.39427784, 0.08)
(4.8079116399999995, 0.09)
(5.2945778, 0.1)
(5.873033339999999, 0.11)
(6.575542799999999, 0.12)
(7.458274480000001, 0.13)
(8.607574600000001, 0.14)
(10.1363477, 0.15)
(12.168198720000001, 0.16)
(14.566627560000002, 0.17)
(16.873253639999998, 0.18)
(18.74669862, 0.19)
(20.199023600000007, 0.2)
(21.38605858, 0.21)
(22.39978192, 0.22)
(23.28544454, 0.23)
(24.075054559999998, 0.24)
(24.800297999999998, 0.25)
(25.468358840000004, 0.26)
(26.097271, 0.27)
(26.69369644, 0.28)
(27.26196842, 0.29)
(27.8057172, 0.3)
(28.32988538, 0.31)
(28.83737736, 0.32)
(29.33490668, 0.33)
(29.825662960000002, 0.34)
(30.308587199999998, 0.35)
(30.78787856, 0.36)
(31.26163452, 0.37)
(31.734244240000002, 0.38)
(32.201864660000005, 0.39)
(32.6697612, 0.4)
(33.13209718, 0.41)
(33.58706964, 0.42)
(34.03575628, 0.43)
(34.47830924, 0.44)
(34.9134825, 0.45)
(35.33925008, 0.46)
(35.757335059999996, 0.47)
(36.16505504, 0.48)
(36.56835602, 0.49)
(36.966124, 0.5)
(37.36004592, 0.51)
(37.753933880000005, 0.52)
(38.146882, 0.53)
(38.54321992, 0.54)
(38.942456, 0.55)
(39.34396464, 0.56)
(39.754073580000004, 0.57)
(40.17044068, 0.58)
(40.599215, 0.59)
(41.0376158, 0.6)
(41.48870056, 0.61)
(41.954981, 0.62)
(42.43719274, 0.63)
(42.93760444, 0.64)
(43.4581625, 0.65)
(43.996672360000005, 0.66)
(44.55508494, 0.67)
(45.13988428, 0.68)
(45.754905619999995, 0.69)
(46.400712, 0.7)
(47.083818, 0.71)
(47.802758239999996, 0.72)
(48.55773354, 0.73)
(49.353498, 0.74)
(50.190161, 0.75)
(51.06836792, 0.76)
(51.98893092, 0.77)
(52.940066, 0.78)
(53.92182310000001, 0.79)
(54.92124140000001, 0.8)
(55.92198776000001, 0.81)
(56.930989079999996, 0.82)
(57.92538734, 0.83)
(58.91393896, 0.84)
(59.8756581, 0.85)
(60.82246432000002, 0.86)
(61.747498600000014, 0.87)
(62.63954124, 0.88)
(63.49528975999999, 0.89)
(64.343268, 0.9)
(65.18548554, 0.91)
(66.03338079999999, 0.92)
(66.9001244, 0.93)
(67.80777035999999, 0.94)
(68.79157319999999, 0.95)
(69.88387807999999, 0.96)
(71.16167318, 0.97)
(72.87253351999999, 0.98)
(75.84571673999999, 0.99)
    };

    \addplot[envoyl4fp, dashed] coordinates {
            (1.505688, 0.0)
    (2.13584998, 0.01)
    (2.382179, 0.02)
    (2.60612898, 0.03)
    (2.83516532, 0.04)
    (3.0692036000000007, 0.05)
    (3.31867396, 0.06)
    (3.58893481, 0.07)
    (3.890928, 0.08)
    (4.234316939999999, 0.09)
    (4.639492800000001, 0.1)
    (5.124172559999998, 0.11)
    (5.72082596, 0.12)
    (6.4787415799999994, 0.13)
    (7.496877200000001, 0.14)
    (8.93199445, 0.15)
    (10.97117584, 0.16)
    (13.49400322, 0.17)
    (15.908091, 0.18)
    (17.86807285, 0.19)
    (19.402142800000014, 0.2)
    (20.61555686, 0.21)
    (21.64434904, 0.22)
    (22.54361209, 0.23)
    (23.34911176, 0.24)
    (24.08524525, 0.25)
    (24.76706764, 0.26)
    (25.405798, 0.27)
    (26.01200248, 0.28)
    (26.590778139999998, 0.29)
    (27.141103899999997, 0.3)
    (27.672861459999996, 0.31)
    (28.186190800000002, 0.32)
    (28.68686378, 0.33)
    (29.17973688, 0.34)
    (29.66712625, 0.35)
    (30.147347279999995, 0.36)
    (30.625290709999998, 0.37)
    (31.10316354, 0.38)
    (31.57616337, 0.39)
    (32.044079, 0.4)
    (32.511526, 0.41)
    (32.97864458, 0.42)
    (33.43731307, 0.43)
    (33.891713, 0.44)
    (34.337764, 0.45)
    (34.77651918, 0.46)
    (35.20706, 0.47)
    (35.62965684, 0.48)
    (36.04492701, 0.49)
    (36.454217, 0.5)
    (36.860249, 0.51)
    (37.26610016, 0.52)
    (37.671009, 0.53)
    (38.07657664, 0.54)
    (38.48760065, 0.55)
    (38.903922, 0.56)
    (39.328552619999996, 0.57)
    (39.76528, 0.58)
    (40.21247291, 0.59)
    (40.67183559999999, 0.6)
    (41.141224629999996, 0.61)
    (41.627949459999996, 0.62)
    (42.12884716000001, 0.63)
    (42.64923024, 0.64)
    (43.193439, 0.65)
    (43.76242778, 0.66)
    (44.357630660000005, 0.67)
    (44.98265088, 0.68)
    (45.63484008, 0.69)
    (46.3176402, 0.7)
    (47.032759, 0.71)
    (47.78582428, 0.72)
    (48.57469013, 0.73)
    (49.40734151999999, 0.74)
    (50.288174, 0.75)
    (51.209895400000015, 0.76)
    (52.159042729999996, 0.77)
    (53.11914422000001, 0.78)
    (54.08604814, 0.79)
    (55.05736140000001, 0.8)
    (56.023136, 0.81)
    (56.97237447999999, 0.82)
    (57.89834189, 0.83)
    (58.81566744, 0.84)
    (59.71522565, 0.85)
    (60.576835519999996, 0.86)
    (61.417299840000005, 0.87)
    (62.24009608, 0.88)
    (63.041535870000004, 0.89)
    (63.828519799999995, 0.9)
    (64.62342753, 0.91)
    (65.42492336000001, 0.92)
    (66.25996433, 0.93)
    (67.15413314, 0.94)
    (68.12655449999997, 0.95)
    (69.22588268, 0.96)
    (70.56591402, 0.97)
    (72.38580334, 0.98)
    (75.63946998999991, 0.99)
        };

    \coordinate (p50 l7fp) at (5.743548, 0.5);
    \coordinate (p50 no-acc) at (36.454217, 0.5);

    \coordinate (p90 l7fp) at (12.259674, 0.9);
    \coordinate (p90 no-acc) at (64.343268, 0.9);

\end{axis}

\node[annot y] (p50 l7fp annot) at ($(p50 l7fp) + (2pt, 0)$) {\scriptsize 5.7};
\node[annot y] (p50 no-acc annot) at ($(p50 no-acc) + (2pt, 0)$) {\scriptsize 37/37};

\node[annot y] (p90 l7fp annot) at ($(p90 l7fp) + (2pt, 0)$) {\scriptsize 12};
\node[annot y, anchor=east] (p90 no-acc annot) at ($(p90 no-acc) - (2pt, 0)$) {\scriptsize 64/64};

\end{tikzpicture}
    \end{subfigure}%
    \begin{subfigure}[t]{0.5\linewidth}
        \input{fig/eval/fp-hr-rate}
    \end{subfigure}
    \caption{\proj reduces the request latency across every percentile while increasing the throughput. }
    \label{fig:eval:fp}
\end{figure}

We use a testing regime that sends TLS-encrypted requests at an increasing rate (closed model) beyond the maximum throughput of the respective application. The regime takes 200\,s and scales the incoming request rate up to 5K\,req/s for the Social Network and the Media Service, and up to 20K\,req/s for the Hotel Reservation. We repeat this experiment 30 times to reduce noise. \Cref{fig:eval:fp} shows the result. On the left, it shows the CDF of the request latency across the entire experiment. On the right, it shows the request completion rate. We found these trends to be consistent with the applications, even if we swapped their respective policies.

\paragraph{Specializing the data plane improves performance} \proj optimizes L7 policies by synthesizing a specialized data plane that is specific to the policy. As we will see in \Cref{sec:eval:cp}, this data plane is, compared to Envoy, more efficient at parsing messages, and enforces the policy without IPC costs. This yields significant performance improvements. Compared to the L4 fast path, \proj reduces the median request latency of the Social Network by $10\times$, the Media Service by $2.5\times$, and the Hotel Reservation by $6\times$. As a result, all applications serve more traffic, too. Compared to the L4 fast path, \proj serves up to 49\%, 38\%, and $3\times$ more traffic. The performance benefits become particularly significant for applications with complex \htwo data planes. \htwo stacks employ expensive state management and coordinate with the peer using control messages. \proj requires little state, and can forward most of the control messages without processing them.

\paragraph{Optimizing IPC alone is not enough} State-of-the-art service meshes improve the performance of L7 policies by optimizing IPC. As \Cref{fig:eval:fp} shows, this yields only minor performance benefits. For the Social Network and Media Service, it serves up to 15\% more traffic, for the Hotel Reservation, which is much more overloaded than the other two applications, the throughput improvement is negligible.

\subsection{What is the Overhead of the Slow Path?}
\label{sec:eval:sp}

Next, we assess the \textit{worst-case scenario}, where all messages are processed on the slow path. In this case, \proj redirects all messages to Envoy for processing. To this end, we use the Echo Service. It exhibits minimal IPC overhead, which makes the overhead of the slow path the most apparent. We configure \proj with the following policy: \medskip

\begin{noindent}
\begin{tabular}{p{\tabcolwidth{0.07}} >{\raggedright}p{\tabcolwidth{0.93}}}
\multicolumn{2}{l}{\textbf{Policy Description}} \tabularnewline
\hline\\
1. & parse HTTP headers with length of $1\,kB$ to $16\,kB$ \tabularnewline
2. & match the HTTP headers against the policy \tabularnewline
3. & route to the service proxy  \tabularnewline
\end{tabular}
\end{noindent}
\medskip

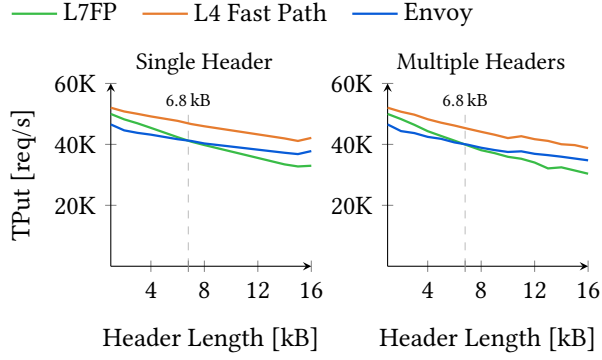
\begin{figure}[t]
    \vspace{-1em}%
    \begin{subfigure}[t]{0.5\linewidth}
        \tikzsetnextfilename{legend-proxy-figure}

\begin{tikzpicture}
    \begin{axis}[
    legend columns = 4,
    legend style={at={(0,0)},draw=none,anchor=south west, /tikz/every even column/.append style={column sep=0.25cm}},
    hide axis,
    xmax=1, ymax=1,
    ]
        \addplot[beeline] coordinates { (0, 0) };
        \addplot[envoyl4fp] coordinates { (0, 0) };
        \addplot[envoy] coordinates { (0, 0) };

        \legend{\proj, L4 Fast Path, Envoy}
    \end{axis}

\end{tikzpicture}
    \end{subfigure} \\
    \begin{subfigure}[b]{0.5\linewidth}
        \tikzsetnextfilename{sp-ssm-rate-figure}

\begin{tikzpicture}
        \begin{axis}[
        ylabel={TPut [req/s]}, xlabel={Header Length [kB]},
        ymin=0, ymax=60000,
        axis lines=left,
        legend columns = 4,
        legend style={at={(-0.125,1.1)},draw=none,anchor=south west, /tikz/every even column/.append style={column sep=0.25cm}},
        y tick label style={
            /pgf/number format/fixed,
            /pgf/number format/precision=1,
            /pgf/number format/1000 sep={},
        },
        yticklabel={\pgfkeys{/pgf/fpu=true}\pgfmathparse{\tick/1000}\pgfmathprintnumber{\pgfmathresult}K},
        ytick={20000, 40000, 60000},
        x tick label style={
            /pgf/number format/fixed,
            /pgf/number format/precision=2,
            /pgf/number format/1000 sep={},
        },
        xticklabel={\pgfkeys{/pgf/fpu=true}\pgfmathparse{\tick/1000}\pgfmathprintnumber{\pgfmathresult}},
        xtick={0, 4000, 8000, 12000, 16000},
        extra x tick label=\empty,
        extra x ticks={6800},
        extra x tick style={
            grid=major,
            grid style=dashed,
        },
        scaled y ticks=false,
        scaled x ticks=false,
        grid=major,
        xmajorgrids=false,
        ymajorgrids=false,
        tick style={
            grid style=dashed,
        },
        xlabel style={anchor=north},
        height=4cm,
        width=\linewidth]

        \addplot[beeline] coordinates {
                    (1000, 49983.81827272674)
        (2000, 48216.17061119228)
        (3000, 46906.53005257152)
        (4000, 45411.104840199856)
        (6000, 42350.798339190944)
        (7000, 40908.42602130549)
        (8000, 39780.436219570496)
        (14000, 33429.91841012961)
        (15000, 32752.19575988184)
        (16000, 32984.30448285619)
                };
        \addplot[envoyl4fp] coordinates {
                    (1000, 52032.38176970309)
        (2000, 50799.208381355704)
        (3000, 50014.98326627209)
        (4000, 49172.46607425381)
        (6000, 47714.04046360355)
        (7000, 46686.098542842316)
        (8000, 45940.14050982145)
        (14000, 41951.10368140231)
        (15000, 41122.397034533475)
        (16000, 42125.948524547464)
                };
        \addplot[envoy] coordinates {
                    (1000, 46543.87409422719)
        (2000, 44599.757946671125)
        (3000, 43776.11045679479)
        (4000, 43188.40415585179)
        (6000, 41691.99877569253)
        (7000, 41129.05053379495)
        (8000, 40295.92964321248)
        (14000, 37246.224384416135)
        (15000, 36807.05967642103)
        (16000, 37800.85556346522)
                };

        \coordinate (top) at (8000,60000);
        \coordinate (thresh) at (6800, 42004.239098227736);

        \end{axis}

        \node[annot x] (annot) at ($(thresh) + (0, 10pt)$) {\scriptsize 6.8\,kB};
        \node[above, text depth=0] at ($(top) + (0, 2pt)$) {\small Single Header};

\end{tikzpicture}%
    \end{subfigure}%
    \begin{subfigure}[b]{0.5\linewidth}
        \tikzsetnextfilename{sp-ssm-rate-figure}

\begin{tikzpicture}
        \begin{axis}[
        xlabel={Header Length [kB]},
        ymin=0, ymax=60000,
        axis lines=left,
        legend columns = 4,
        legend style={at={(-0.125,1.1)},draw=none,anchor=south west, /tikz/every even column/.append style={column sep=0.25cm}},
        y tick label style={
            /pgf/number format/fixed,
            /pgf/number format/precision=1,
            /pgf/number format/1000 sep={},
        },
        yticklabel={\pgfkeys{/pgf/fpu=true}\pgfmathparse{\tick/1000}\pgfmathprintnumber{\pgfmathresult}K},
        ytick={20000, 40000, 60000},
        x tick label style={
            /pgf/number format/fixed,
            /pgf/number format/precision=2,
            /pgf/number format/1000 sep={},
        },
        xticklabel={\pgfkeys{/pgf/fpu=true}\pgfmathparse{\tick/1000}\pgfmathprintnumber{\pgfmathresult}},
        xtick={0, 4000, 8000, 12000, 16000},
        extra x tick label=\empty,
        extra x ticks={6800},
        extra x tick style={
            grid=major,
            grid style=dashed,
        },
        scaled y ticks=false,
        scaled x ticks=false,
        grid=major,
        xmajorgrids=false,
        ymajorgrids=false,
        tick style={
            grid style=dashed,
        },
        xlabel style={anchor=north},
        height=4cm,
        width=\linewidth]

        \addplot[beeline] coordinates {
                    (1000, 49983.81827272674)
        (2000, 48307.966796239576)
        (3000, 46421.016744804016)
        (4000, 44332.85504484861)
        (5000, 42761.31024841055)
        (6000, 41228.8414180447)
        (7000, 39625.81225763991)
        (8000, 38086.361072631385)
        (9000, 37130.64419692924)
        (10000, 35919.40601518199)
        (11000, 35278.63024102421)
        (12000, 34039.91144838366)
        (13000, 32110.2063468339)
        (14000, 32463.64139965084)
        (15000, 31424.84745160119)
        (16000, 30370.597429529917)
                };
        \addplot[envoyl4fp] coordinates {
                    (1000, 52032.38176970309)
        (2000, 50717.59831366265)
        (3000, 49748.59677300355)
        (4000, 48251.27952090322)
        (5000, 47119.500699672215)
        (6000, 46181.45058454017)
        (7000, 45129.84279863345)
        (8000, 44122.14356458416)
        (9000, 43168.29288846421)
        (10000, 42061.3508143521)
        (11000, 42678.356938110184)
        (12000, 41704.10997889472)
        (13000, 41136.719643617)
        (14000, 40039.55082386096)
        (15000, 39757.703747649255)
        (16000, 38773.2311079476)
                };
        \addplot[envoy] coordinates {
                    (1000, 46543.87409422719)
        (2000, 44358.7147375906)
        (3000, 43706.85658899515)
        (4000, 42461.29788266931)
        (5000, 41819.58515391234)
        (6000, 40647.29920805249)
        (7000, 39887.78536412989)
        (8000, 38893.85155317816)
        (9000, 38133.07333192839)
        (10000, 37506.52504410881)
        (11000, 37718.14579333522)
        (12000, 36896.37359104415)
        (13000, 36463.92986512298)
        (14000, 35952.7966329449)
        (15000, 35354.761189436176)
        (16000, 34762.367847743386)
                };

        \coordinate (top) at (8000,60000);
        \coordinate (thresh) at (6800, 42004.239098227736);

        \end{axis}

        \node[annot x] (annot) at ($(thresh) + (0, 10pt)$) {\scriptsize 6.8\,kB};
        \node[above, text depth=0] at ($(top) + (0, 2pt)$) {\small Multiple Headers};

\end{tikzpicture}%
    \end{subfigure}
    \caption{\proj's slow path improves the throughput for messages smaller than 6.8\,kB. For larger headers, the parsing overhead outweighs the IPC optimizations.}
    \label{fig:eval:sp:ssm}
\end{figure}

In this experiment, we scale the length of the HTTP headers from 1\,kB to 16\,kB. We increment the header length by 1\,kB at a time and repeat the experiment with two different HTTP header compositions. In the first composition, \emph{Single Header}, we fix the number of parsed and matched headers to one. In the second composition, \emph{Multiple Headers}, we fix the size of each HTTP header to 1\,kB. The load generator establishes 3000 connections and sends as many messages as possible in one minute. \Cref{fig:eval:sp:ssm} summarizes the results.

\paragraph{\proj accelerates the slow path too} The number of HTTP headers has a negligible impact on the performance of \proj. In both configurations, \proj remains more performant than Envoy for HTTP headers that are smaller than 6.8\,kB. However, as the size of the HTTP headers increases, the throughput decreases, e.g., by 11\% for 10\,kB headers.

In this experiment, \proj's runtime is dominated by its parsing overhead, which is a function of the number of parsed bytes. For a 1\,kB HTTP header, 73\% of the runtime is spent on parsing. This overhead is fundamental, as \proj must iterate over the full HTTP header to match the message against the configured policy. The remainder of the runtime is implementation-specific and accounts for data copies, string comparisons and eBPF map queries (see \Cref{sec:eval:cp}).

We want to emphasize that this experiment pushes the ratio between IPC cost and parsing overhead to the extreme. In this experiment, IPC cost is minor with only two pods that process the request. On the other hand, the request sizes are large in comparison to cloud-scale deployments. \cite{seemakhupt_cloud} reports that half of the observed RPC calls in production have a median request size below 1.5\,kB, with responses blow 315\,B. For requests of this size, \proj achieves roughly 8\% more throughput than Envoy. Finally, it goes without saying that \proj should not be deployed in a service mesh where it cannot process any message directly in the kernel. Despite this, \proj can efficiently record traffic telemetry. Future versions can use this data to dynamically (de)active the fast path on a per-connection basis. Thus, we expect \proj to be able to accelerate a vast majority of workloads, even if in some cases it has to resort to the slow path.


\subsection{How does the Fast Path Scale?}
\label{sec:eval:cp}

The previous experiments show that \proj's fast path accelerates realistic workloads significantly, and the slow path only induces a slowdown in extreme cases. In this experiment, we validate whether the benefits of the fast path also hold as policies become increasingly complex. As in \Cref{sec:eval:sp}, we deploy the Echo Service to minimize IPC overhead, and consequently limit \proj to rely more on its parsing and processing pipelines for performance gains. We configure the service proxy with four different policies, each of which can be tuned to be more {\em complex} via a tuneable parameter {\em c}: \medskip

\begin{noindent}
\begin{tabular}{p{\tabcolwidth{0.2}}p{\tabcolwidth{0.12}} >{\raggedright}p{\tabcolwidth{0.68}}}
\textbf{Policy} & \multicolumn{2}{l}{\textbf{Description}} \tabularnewline
\hline\\
Route & 1.   & parse the header with length~$c \cdot 16\,kB$ \tabularnewline
        & 2.   & match one HTTP header with length~$c \cdot 16\,kB$ \tabularnewline
        & 3.   & record traffic telemetry \tabularnewline\tabularnewline
RBAC & 1-3. & enforce the Route policy \tabularnewline
        & 4.   & enforce $c \cdot 100$ RBAC policies \tabularnewline\tabularnewline
JWT & 1-4. & enforce the RBAC policy \tabularnewline
        & 5.   & authenticate the signature of a JWT with total length $c \cdot 3\,kB$ \tabularnewline\tabularnewline
Mutate & 1-5. & enforce the JWT policy \tabularnewline
        & 6.   & remove $c \cdot 16\,kB$ bytes from the headers \tabularnewline
\end{tabular}
\end{noindent}
\medskip

These policies are designed to scale the complexity along two dimensions: the number of actions and the complexity of the individual parsers and actions. To build increasingly complex policies, we can nest actions recursively. As we can see from the policy description, the complexity parameter {\em c} affects multiple policy components simultaneously. For example, RBAC$[c=1]$ parses one 16\,kB-sized HTTP header \emph{and} enforces 100 RBAC rules. The final policy, \emph{Mutate}, represents the most complex policy that the current implementation of \proj can offload (see \Cref{sec:implementation} for a discussion on how to alleviate this limit). Given that the parsing complexity is a function of the total number of bytes parsed and remains largely independent of the number of HTTP headers (\Cref{sec:eval:sp}), this experiment uses only one header.

\begin{figure}[t]
    \vspace{-1em}%
    \begin{subfigure}[t]{0.5\linewidth}
        \tikzsetnextfilename{legend-proxy-figure}

\begin{tikzpicture}
    \begin{axis}[
    legend columns = 4,
    legend style={at={(0,0)},draw=none,anchor=south west, /tikz/every even column/.append style={column sep=0.25cm}},
    hide axis,
    xmax=1, ymax=1,
    ]
        \addplot[beeline] coordinates { (0, 0) };
        \addplot[envoyl4fp] coordinates { (0, 0) };
        \addplot[envoy] coordinates { (0, 0) };

        \legend{\proj, L4 Fast Path, Envoy}
    \end{axis}

\end{tikzpicture}
    \end{subfigure} \\
    \begin{subfigure}[b]{0.6\linewidth}
        \tikzsetnextfilename{pc-ssm-p1-rate-figure}

\begin{tikzpicture}
        \begin{axis}[
        xlabel={Complexity},
        ylabel={TPut [req/s]},
        ymin=0, ymax=100000,
        axis lines=left,
        legend columns = 4,
        legend style={at={(0,1.1)},draw=none,anchor=south west, /tikz/every even column/.append style={column sep=0.25cm}},
        y tick label style={
            /pgf/number format/fixed,
            /pgf/number format/precision=1,
            /pgf/number format/1000 sep={},
        },
        yticklabel={\pgfkeys{/pgf/fpu=true}\pgfmathparse{\tick/1000}\pgfmathprintnumber{\pgfmathresult}K},
        ytick={0, 25000, 50000, 75000, 100000},
        x tick label style={
            /pgf/number format/fixed,
            /pgf/number format/precision=2,
            /pgf/number format/1000 sep={},
        },
        xticklabel={\pgfkeys{/pgf/fpu=true}\pgfmathparse{\tick/16000}\pgfmathprintnumber{\pgfmathresult}},
        xtick={0, 4000, 8000, 12000, 16000},
        extra x tick label=\empty,
        extra x ticks={8000, 16000},
        extra x tick style={
            grid=major,
            grid style=dashed,
        },
        scaled y ticks=false,
        scaled x ticks=false,
        grid=major,
        ymajorgrids=false,
        xmajorgrids=false,
        tick style={
            grid style=dashed,
        },
        xlabel style={anchor=north},
        height=4cm,
        width=4.5cm]

        \addplot[beeline] coordinates {
                    (1000, 82710.05605564518)
        (2000, 88814.08707348001)
        (3000, 85826.03275546087)
        (4000, 82143.85439646614)
        (6000, 77634.4672728244)
        (7000, 72952.59601356406)
        (8000, 72260.17809203493)
        (14000, 60577.90173514409)
        (15000, 58729.85845181104)
        (16000, 57166.362858688124)
                };
        \addplot[envoyl4fp] coordinates {
                    (1000, 51454.219871375215)
        (2000, 50776.13711186833)
        (3000, 50332.561500120726)
        (4000, 49016.17446270528)
        (6000, 47700.57020019045)
        (7000, 46981.68135956014)
        (8000, 45881.71822380262)
        (14000, 41872.36915666412)
        (15000, 41307.258972674914)
        (16000, 40607.15461692194)
                };
        \addplot[envoy] coordinates {
                    (1000, 46710.35784442914)
        (2000, 44586.892757736234)
        (3000, 43956.994903827115)
        (4000, 43028.660179853694)
        (6000, 41843.4852383186)
        (7000, 41327.38072273717)
        (8000, 40244.65950544221)
        (14000, 37144.992047076994)
        (15000, 36674.83527856715)
        (16000, 36211.23677319304)
                };

        \coordinate (top) at (8000,100000);
        \coordinate (simple) at (8000, 72260);
        \coordinate (hard) at (16000, 57166);

        \end{axis}

        \node[annot x, text height=20pt, minimum height=20pt] (annot simple) at ($(simple) + (0, 4pt)$) {\scriptsize +57\%};
        \node[annot x, text height=40pt, minimum height=40pt] (annot hard) at ($(hard) + (0, 4pt)$) {\scriptsize +40\%};
        \node[above, text depth=0] at ($(top) + (0, 2pt)$) {\small Route};

        \end{tikzpicture}%
    \end{subfigure}%
    \begin{subfigure}[b]{0.4\linewidth}
        \tikzsetnextfilename{pc-ssm-p2-rate-figure}

\begin{tikzpicture}
        \begin{axis}[
        xlabel={Complexity},
        ymin=0, ymax=100000,
        axis lines=left,
        legend columns = 4,
        legend style={at={(0,1.1)},draw=none,anchor=south west, /tikz/every even column/.append style={column sep=0.25cm}},
        y tick label style={
            /pgf/number format/fixed,
            /pgf/number format/precision=1,
            /pgf/number format/1000 sep={},
        },
        yticklabel=\empty,
        ytick={0, 25000, 50000, 75000, 100000},
        x tick label style={
            /pgf/number format/fixed,
            /pgf/number format/precision=2,
            /pgf/number format/1000 sep={},
        },
        xticklabel={\pgfkeys{/pgf/fpu=true}\pgfmathparse{\tick/16000}\pgfmathprintnumber{\pgfmathresult}},
        xtick={0, 4000, 8000, 12000, 16000},
        extra x tick label=\empty,
        extra x ticks={8000, 16000},
        extra x tick style={
            grid=major,
            grid style=dashed,
        },
        scaled y ticks=false,
        scaled x ticks=false,
        grid=major,
        ymajorgrids=false,
        xmajorgrids=false,
        tick style={
            grid style=dashed,
        },
        xlabel style={anchor=north},
        height=4cm,
        width=4.5cm]

        \addplot[beeline] coordinates {
                    (1000, 91141.2787960349)
        (2000, 78906.21688213637)
        (3000, 81344.57535144116)
        (4000, 81743.19840070116)
        (6000, 76545.50409531799)
        (7000, 70508.59226696978)
        (8000, 71842.09667093927)
        (14000, 59756.515243411144)
        (15000, 58604.816986862905)
        (16000, 56662.54068148427)
                };
        \addplot[envoyl4fp] coordinates {
                    (1000, 53045.512955756116)
        (2000, 52774.9541028429)
        (3000, 51585.54087185694)
        (4000, 49723.865369759194)
        (6000, 48294.778041812046)
        (7000, 47288.73569973512)
        (8000, 46111.33895921248)
        (14000, 42157.37796775849)
        (15000, 41752.22606398748)
        (16000, 40711.341043166)
                };

        \addplot[envoy] coordinates {
                    (1000, 47151.755705502124)
        (2000, 45935.800201818405)
        (3000, 45080.15180760356)
        (4000, 44783.85324705084)
        (6000, 43043.01978836819)
        (7000, 42047.8350776514)
        (8000, 40981.06406599459)
        (14000, 37646.756036095365)
        (15000, 37011.97936938353)
        (16000, 36515.2395669046)
                };

        \coordinate (top) at (8000,100000);
        \coordinate (simple) at (8000, 71842);
        \coordinate (hard) at (16000, 56662);

        \end{axis}

        \node[annot x, text height=20pt, minimum height=20pt] (annot simple) at ($(simple) + (0, 4pt)$) {\scriptsize +55\%};
        \node[annot x, text height=40pt, minimum height=40pt] (annot hard) at ($(hard) + (0, 4pt)$) {\scriptsize +39\%};
        \node[above, text depth=0] at ($(top) + (0, 2pt)$) {\small RBAC};

        \end{tikzpicture}%
    \end{subfigure} \\
    \begin{subfigure}[b]{0.6\linewidth}
        \tikzsetnextfilename{pc-ssm-p3-rate-figure}

\begin{tikzpicture}
        \begin{axis}[
        ylabel={TPut [req/s]},
        xlabel={Complexity},
        ymin=0, ymax=100000,
        axis lines=left,
        legend columns = 4,
        legend style={at={(0,1.1)},draw=none,anchor=south west, /tikz/every even column/.append style={column sep=0.25cm}},
        y tick label style={
            /pgf/number format/fixed,
            /pgf/number format/precision=1,
            /pgf/number format/1000 sep={},
        },
        yticklabel={\pgfkeys{/pgf/fpu=true}\pgfmathparse{\tick/1000}\pgfmathprintnumber{\pgfmathresult}K},
        ytick={0, 25000, 50000, 75000, 100000},
        x tick label style={
            /pgf/number format/fixed,
            /pgf/number format/precision=2,
            /pgf/number format/1000 sep={},
        },
        xticklabel={\pgfkeys{/pgf/fpu=true}\pgfmathparse{\tick/16000}\pgfmathprintnumber{\pgfmathresult}},
        xtick={0, 4000, 8000, 12000, 16000},
        extra x tick label=\empty,
        extra x ticks={8000, 16000},
        extra x tick style={
            grid=major,
            grid style=dashed,
        },
        scaled y ticks=false,
        scaled x ticks=false,
        grid=major,
        ymajorgrids=false,
        xmajorgrids=false,
        tick style={
            grid style=dashed,
        },
        xlabel style={anchor=north},
        height=4cm,
        width=4.5cm]

        \addplot[beeline] coordinates {
                    (1000, 85909.8313469897)
        (2000, 78230.12477784093)
        (3000, 77116.08745877529)
        (5000, 69246.09629572504)
        (6000, 66175.45122794466)
        (7000, 63309.59927802975)
        (8000, 60974.64763884088)
        (10000, 55546.581189653676)
        (11000, 53412.94548450672)
        (12000, 51169.03334076748)
        (13000, 49917.916132498474)
        (14000, 47548.138055427946)
        (15000, 46067.7650856236)
        (16000, 44922.92706160679)
                };
        \addplot[envoyl4fp] coordinates {
                    (1000, 47287.8564138719)
        (2000, 45949.67238706436)
        (3000, 44681.414851812624)
        (4000, 42912.139670460325)
        (5000, 41633.68357597639)
        (6000, 40366.6189890599)
        (7000, 39166.985890049764)
        (8000, 38284.80348380905)
        (10000, 36244.14997810915)
        (11000, 35358.08272901362)
        (12000, 34369.54901499446)
        (13000, 33652.76963390203)
        (14000, 32715.67756092833)
        (15000, 32010.003229867798)
        (16000, 31097.296948276595)
                };
        \addplot[envoy] coordinates {
                    (1000, 42694.82956385025)
        (2000, 41030.087516802254)
        (3000, 40053.36943534341)
        (4000, 38742.57474007692)
        (5000, 37567.92941247452)
        (6000, 36502.3300451589)
        (7000, 35499.38311289083)
        (8000, 34452.427464760134)
        (10000, 32888.557520599206)
        (11000, 31871.854375501232)
        (12000, 31166.29388417362)
        (13000, 30488.03053600688)
        (14000, 29808.93216382371)
        (15000, 29134.72534405653)
        (16000, 28397.226128557064)
                };

        \coordinate (top) at (8000,100000);
        \coordinate (simple) at (8000, 60974);
        \coordinate (hard) at (16000, 44922);

        \end{axis}

        \node[annot x, text height=20pt, minimum height=20pt] (annot simple) at ($(simple) + (0, 5pt)$) {\scriptsize +59\%};
        \node[annot x, text height=40pt, minimum height=40pt] (annot hard) at ($(hard) + (0, 4pt)$) {\scriptsize +44\%};
        \node[above, text depth=0] at ($(top) + (0, 2pt)$) {\small JWT};

        \end{tikzpicture}%
    \end{subfigure}%
    \begin{subfigure}[b]{0.4\linewidth}
        \tikzsetnextfilename{pc-ssm-p4-rate-figure}

\begin{tikzpicture}
        \begin{axis}[
        xlabel={Complexity},
        ymin=0, ymax=100000,
        axis lines=left,
        legend columns = 4,
        legend style={at={(0,1.1)},draw=none,anchor=south west, /tikz/every even column/.append style={column sep=0.25cm}},
        y tick label style={
            /pgf/number format/fixed,
            /pgf/number format/precision=1,
            /pgf/number format/1000 sep={},
        },
        yticklabel=\empty,
        ytick={0, 25000, 50000, 75000, 100000},
        x tick label style={
            /pgf/number format/fixed,
            /pgf/number format/precision=2,
            /pgf/number format/1000 sep={},
        },
        xticklabel={\pgfkeys{/pgf/fpu=true}\pgfmathparse{\tick/16000}\pgfmathprintnumber{\pgfmathresult}},
        xtick={0, 4000, 8000, 12000, 16000},
        extra x tick label=\empty,
        extra x ticks={8000, 16000},
        extra x tick style={
            grid=major,
            grid style=dashed,
        },
        scaled y ticks=false,
        scaled x ticks=false,
        grid=major,
        ymajorgrids=false,
        xmajorgrids=false,
        tick style={
            grid style=dashed,
        },
        xlabel style={anchor=north},
        height=4cm,
        width=4.5cm]

        \addplot[beeline] coordinates {
                    (1000, 85525.10405990525)
        (2000, 81501.79870531727)
        (3000, 77479.91952650114)
        (4000, 68814.09588871546)
        (5000, 70103.25522688396)
        (6000, 67515.38819143621)
        (7000, 64376.93681499335)
        (8000, 62333.13788277094)
        (10000, 57135.99345678873)
        (11000, 54565.02726672691)
        (12000, 53209.80450146824)
        (13000, 51115.905765251424)
        (14000, 49468.258386598274)
        (15000, 47541.798748311856)
        (16000, 46361.08027317366)
                };
        \addplot[envoyl4fp] coordinates {
                    (1000, 44571.812400284056)
        (2000, 43629.43299104161)
        (3000, 42032.63677996957)
        (4000, 40691.675271855056)
        (5000, 39914.18333310847)
        (6000, 38563.72619945843)
        (7000, 37742.18883711214)
        (8000, 35941.906993477016)
        (9000, 39266.830039718014)
        (10000, 34981.903062902464)
        (11000, 34061.48684009679)
        (12000, 32802.17051774355)
        (13000, 32571.08317758496)
        (14000, 31693.78875573894)
        (15000, 30979.536172348955)
        (16000, 30424.693644487128)
                };
        \addplot[envoy] coordinates {
                    (1000, 40156.16829537245)
        (2000, 39180.541718048786)
        (3000, 37929.65000112507)
        (4000, 37044.89475116151)
        (5000, 36436.47730707785)
        (6000, 35346.94943674958)
        (7000, 34354.22882733383)
        (8000, 32853.00589851526)
        (9000, 35608.394455696536)
        (10000, 31804.32330905851)
        (11000, 31537.90309949598)
        (12000, 30575.68348021743)
        (13000, 30272.675051816255)
        (14000, 29212.194454244236)
        (15000, 28339.412941462724)
        (16000, 28051.82564236411)
                };

        \coordinate (top) at (8000,100000);
        \coordinate (simple) at (8000, 62333);
        \coordinate (hard) at (16000, 46361);

        \end{axis}

        \node[annot x, text height=20pt, minimum height=20pt] (annot simple) at ($(simple) + (0, 4pt)$) {\scriptsize +73\%};
        \node[annot x, text height=40pt, minimum height=40pt] (annot hard) at ($(hard) + (0, 4pt)$) {\scriptsize +52\%};
        \node[above, text depth=0] at ($(top) + (0, 2pt)$) {\small Mutate};

        \end{tikzpicture}%
    \end{subfigure}
    \caption{Even for the most complex policies, \proj improves the throughput of the L4 fast path by at least 39\%. For above-average complex policies, the throughput improvement can reach up to 73\%.}
    \label{fig:eval:pc:ssm}
\end{figure}
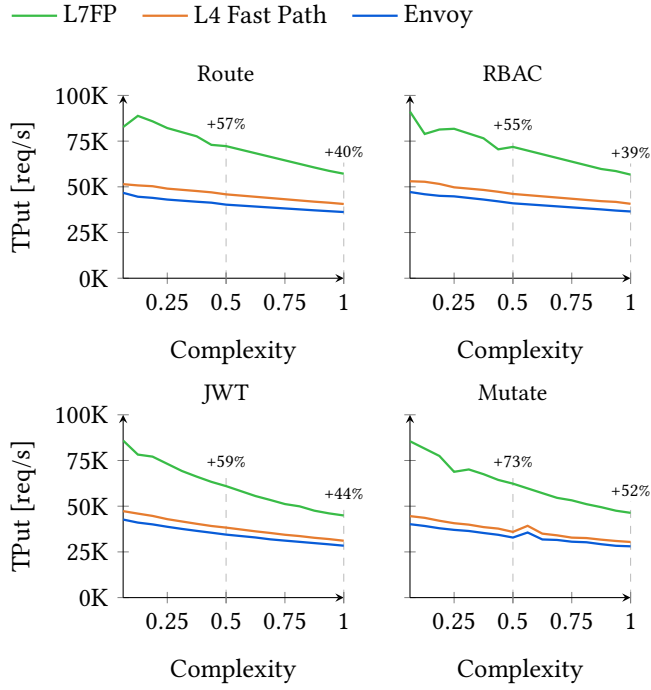

\paragraph{\proj's performance is a function of policy complexity} As shown in \Cref{fig:eval:pc:ssm}, we observe the same trend for all four policies. In general, \proj consistently outperforms both configurations of Envoy. However, this performance gain diminishes as policies become more complex.
For example, consider the policy JWT$[c=0.5]$, which parses and matches an 8\,kB long HTTP header, enforces 50 RBAC policies, and authenticates a 1.5\,kB long JWT. With this configuration, \proj achieves a 59\% higher throughput than the L4 fast path. For JWT$[c=1]$, this improvement shrinks to 44\%.

We note that our per-node deployment is an extreme case. Typically, the service proxy's overhead is multiplied in a per-pod deployment because more than two instances process each request. Additionally, as mentioned earlier, \cite{seemakhupt_cloud} reports that half of all RPC calls in their data center have a median request size below 1.5\,kB in their production systems, which corresponds to policies with $c\leq0.1$. To summarize, for all policies that the current implementation of \proj is able to offload, it is more efficient than Envoy, even when accelerated with an L4 fast path.

\paragraph{Understanding \proj's overheads} The previous experiment shows that the performance of \proj degrades more quickly than Envoy's. This is to be expected, as eBPF programs are executed in a virtual machine that lacks some runtime optimizations like SIMD~\cite{miano_fast, shahinfar_demystifying}. Moreover, calls to a \texttt{bpf\_helper} function can be expensive, hurting performance further. Thus, the more complex a policy is, the smaller the performance gain of \proj becomes.

\begin{figure}[t]
    \vspace{-1em}%
    \begin{subfigure}[t]{0.5\linewidth}
        \tikzsetnextfilename{legend-dissect-figure}

\begin{tikzpicture}
    \begin{axis}[
    ybar stacked,
    legend columns = 4,
    legend cell align={left},
    legend style={at={(0,0)},draw=none,anchor=south west, /tikz/every even column/.append style={column sep=0.25cm}},
    hide axis,
    xmax=1, ymax=1,
    ]
        \addplot[fill=uchu-pink-1, draw=uchu-pink-6] coordinates { (0, 0) };
        \addplot[fill=uchu-red-1, draw=uchu-red-6] coordinates { (0, 0) };
        \addplot[fill=uchu-purple-1, draw=uchu-purple-6] coordinates { (0, 0) };
        \addplot[pattern=north west lines, draw=uchu-gray-6, pattern color=uchu-gray-6] coordinates { (0, 0) };

        \legend{Parsing, Policy Enforcement, IPC, Other}
    \end{axis}

\end{tikzpicture}
    \end{subfigure}\\
    \begin{subfigure}[b]{0.4\linewidth}
        \tikzsetnextfilename{pc-ssm-dissect-beeline-figure}

\begin{tikzpicture}
    \begin{axis}[
    xlabel={\textcolor{white}{Complexity}},
    ylabel={Overhead [ms]},
    ymin=0, ymax=1,
    axis lines=left,
    cycle list name=uchu,
    x tick label style={
        /pgf/number format/fixed,
        /pgf/number format/precision=2,
        /pgf/number format/1000 sep={},
    },
    xticklabel={\pgfkeys{/pgf/fpu=true}\pgfmathparse{\tick/16000}\pgfmathprintnumber{\pgfmathresult}},
    xtick={0, 8000, 16000},
    ytick={0, 0.5, 1},
    scaled y ticks=false,
    scaled x ticks=false,
    grid=major,
    ymajorgrids=true,
    xmajorgrids=false,
    tick style={
        grid style=dashed,
    },
    xlabel style={anchor=north},
    enlarge x limits={abs=4pt,upper},
    height=4cm,
    width=3.5cm]

    \path[name path=axis] (axis cs:0,0) -- (axis cs:16000,0);
    \addplot[name path=Parsing, uchu-pink-6] coordinates {(1000, 0.015977712381893532) (2000, 0.025748671122355755) (3000, 0.03537227484493036) (4000, 0.04612025285759387) (5000, 0.05067377639625301) (6000, 0.0649461986204063) (7000, 0.07395853761670244) (8000, 0.09368608081522327) (10000, 0.10707148386556166) (11000, 0.11996506090017699) (12000, 0.12915077641140302) (13000, 0.13142464571465023) (14000, 0.1534160764129228) (15000, 0.1538133424354802) (16000, 0.1728656328690344)};
    \addplot[uchu-pink-1] fill between[of = Parsing and axis];
    \addplot[name path=Processing, uchu-red-6] coordinates {(1000, 0.03498232513647318) (2000, 0.050620624963103523) (3000, 0.06796673851857538) (4000, 0.08698223127481339) (5000, 0.09950303000311383) (6000, 0.12094623468986171) (7000, 0.13676875109174555) (8000, 0.16628762260028113) (10000, 0.19315198374716855) (11000, 0.21456605368107282) (12000, 0.2300761479901353) (13000, 0.2376275371513028) (14000, 0.26910465651215415) (15000, 0.2763175006346563) (16000, 0.30680250067520304)};
    \addplot[uchu-red-1] fill between[of = Processing and Parsing];

    \coordinate (top) at (8000,1);
    \coordinate (hard) at (16000, 0.30680250067520304);

    \end{axis}

    \node[annot x, minimum height=0, text height=0, anchor=south east] (annot) at ($(hard) + (0, 2pt)$) {\scriptsize +877\%};
    \node[above, text depth=0] at ($(top) + (0, 2pt)$) {\small \proj};

\end{tikzpicture}
    \end{subfigure}%
    \begin{subfigure}[b]{0.3\linewidth}
        \tikzsetnextfilename{pc-ssm-dissect-l4fp-figure}

\begin{tikzpicture}
    \begin{axis}[
    xlabel={Complexity},
    ymin=0, ymax=1,
    axis lines=left,
    cycle list name=uchu,
    legend columns = 4,
    legend style={at={(0,1.1)},draw=none,anchor=south west, /tikz/every even column/.append style={column sep=0.25cm}},
    x tick label style={
        /pgf/number format/fixed,
        /pgf/number format/precision=2,
        /pgf/number format/1000 sep={},
    },
    xticklabel={\pgfkeys{/pgf/fpu=true}\pgfmathparse{\tick/16000}\pgfmathprintnumber{\pgfmathresult}},
    xtick={0, 8000, 16000},
    ytick={0, 0.5, 1},
    yticklabel=\empty,
    scaled y ticks=false,
    scaled x ticks=false,
    grid=major,
    ymajorgrids=true,
    xmajorgrids=false,
    tick style={
        grid style=dashed,
    },
    xlabel style={anchor=north},
    enlarge x limits={abs=4pt,upper},
    height=4cm,
    width=3.5cm]

    \path[name path=axis] (axis cs:0,0) -- (axis cs:16000,0);
    \addplot[name path=Parsing, uchu-pink-6] coordinates {(1000, 0.19269193121224604) (2000, 0.21832926371539998) (3000, 0.23123766895414588) (4000, 0.24737700743748844) (5000, 0.2647388000633597) (6000, 0.28087970263248097) (7000, 0.2943794214215513) (8000, 0.3042528210272179) (10000, 0.3452809265013542) (11000, 0.3669871578670242) (12000, 0.38124743365549185) (13000, 0.39521919565273805) (14000, 0.4086911600051032) (15000, 0.4318009106524454) (16000, 0.4408332093215611)};
    \addplot[uchu-pink-1] fill between[of = Parsing and axis];
    \addplot[name path=Processing, uchu-red-6] coordinates {(1000, 0.20593692336061187) (2000, 0.23234189248679032) (3000, 0.24511259696396331) (4000, 0.2617538132724408) (5000, 0.27946081381111887) (6000, 0.2954194261500818) (7000, 0.30903283071410576) (8000, 0.3187942426895274) (10000, 0.36046752048782865) (11000, 0.38268137643001177) (12000, 0.3972772198177763) (13000, 0.41133872101753266) (14000, 0.42561860414549335) (15000, 0.4494067221282186) (16000, 0.45766119265359695)};
    \addplot[uchu-red-1] fill between[of = Processing and Parsing];
    \addplot[name path=IPC, uchu-purple-6] coordinates {(1000, 0.21882277162145652) (2000, 0.2456831545695655) (3000, 0.25856942213993017) (4000, 0.27514436000298403) (5000, 0.2930235945832856) (6000, 0.308724875038519) (7000, 0.3221922082991401) (8000, 0.3315434231046182) (10000, 0.3736556823459453) (11000, 0.39617327441304784) (12000, 0.41076502819202365) (13000, 0.4250014814045566) (14000, 0.43884041004726954) (15000, 0.4630405008946865) (16000, 0.47061617314745136)};
    \addplot[uchu-purple-1] fill between[of = IPC and Processing];
    \addplot[name path=Other, uchu-gray-6] coordinates {(1000, 0.27199102332618735) (2000, 0.35167559832920364) (3000, 0.37919972581587336) (4000, 0.41368837406406433) (5000, 0.39158302420906116) (6000, 0.40323619398798355) (7000, 0.4424841004080976) (8000, 0.5247619928248337) (10000, 0.5053979374806558) (11000, 0.5301517025564284) (12000, 0.5541185520677294) (13000, 0.5499071160115148) (14000, 0.642832815769337) (15000, 0.6452139498470962) (16000, 0.6835501569066822)};
    \addplot[pattern=north west lines, draw=uchu-gray-6, pattern color=uchu-gray-6] fill between[of = Other and IPC];

    \coordinate (top) at (8000,1);
    \coordinate (hard) at (16000, 0.6835501569066822);

    \end{axis}

    \node[annot x, minimum height=0, text height=0, anchor=south east] (annot) at ($(hard) + (0, 2pt)$) {\scriptsize +251\%};
    \node[above, text depth=0] at ($(top) + (0, 2pt)$) {\small L4 Fast Path};

\end{tikzpicture}
    \end{subfigure}%
    \begin{subfigure}[b]{0.3\linewidth}
        \tikzsetnextfilename{pc-ssm-dissect-l4fp-figure}

\begin{tikzpicture}
    \begin{axis}[
    xlabel={\textcolor{white}{Complexity}},
    ymin=0, ymax=1,
    axis lines=left,
    cycle list name=uchu,
    legend columns = 4,
    legend style={at={(0,1.1)},draw=none,anchor=south west, /tikz/every even column/.append style={column sep=0.25cm}},
    x tick label style={
        /pgf/number format/fixed,
        /pgf/number format/precision=2,
        /pgf/number format/1000 sep={},
    },
    xticklabel={\pgfkeys{/pgf/fpu=true}\pgfmathparse{\tick/16000}\pgfmathprintnumber{\pgfmathresult}},
    xtick={0, 8000, 16000},
    ytick={0, 0.5, 1},
    yticklabel=\empty,
    scaled y ticks=false,
    scaled x ticks=false,
    grid=major,
    ymajorgrids=true,
    xmajorgrids=false,
    tick style={
        grid style=dashed,
    },
    xlabel style={anchor=north},
    enlarge x limits={abs=4pt,upper},
    height=4cm,
    width=3.5cm]

    \path[name path=axis] (axis cs:0,0) -- (axis cs:16000,0);
    \addplot[name path=Parsing, uchu-pink-6] coordinates {(1000, 0.19688122556960028) (2000, 0.2210110166330776) (3000, 0.23594121070488322) (4000, 0.2535267591417497) (5000, 0.2700252126193259) (6000, 0.28534881398353695) (7000, 0.3000767500921748) (8000, 0.3182453870918914) (10000, 0.34521873742162673) (11000, 0.36483623718270997) (12000, 0.3812545724351863) (13000, 0.39827992368581244) (15000, 0.42916589804449035) (16000, 0.4474084968735978)};
    \addplot[uchu-pink-1] fill between[of = Parsing and axis];
    \addplot[name path=Processing, uchu-red-6] coordinates {(1000, 0.2481650719999827) (2000, 0.2746989279081937) (3000, 0.2897557911331434) (4000, 0.30789073612100704) (5000, 0.32423621540454306) (6000, 0.34922080214065065) (7000, 0.36876938408949117) (8000, 0.3876459586077303) (10000, 0.41419113262284285) (11000, 0.4341932747271876) (12000, 0.45043974171624057) (13000, 0.4687200592602123) (15000, 0.49980260502079205) (16000, 0.5187829709092678)};
    \addplot[uchu-red-1] fill between[of = Processing and Parsing];
    \addplot[name path=IPC, uchu-purple-6] coordinates {(1000, 0.2774153040610745) (2000, 0.30525574015032914) (3000, 0.32004419585470945) (4000, 0.33808550106983964) (5000, 0.354555248574144) (6000, 0.378650101221916) (7000, 0.3979665383539145) (8000, 0.416923852318125) (10000, 0.4433258910521234) (11000, 0.4636248234225643) (12000, 0.4794845509329839) (13000, 0.4980574694811366) (15000, 0.5291006340398801) (16000, 0.5480218248793067)};
    \addplot[uchu-purple-1] fill between[of = IPC and Processing];
    \addplot[name path=Other, uchu-gray-6] coordinates {(1000, 0.3816598402107155) (2000, 0.4123255316779147) (3000, 0.4805278597402678) (4000, 0.45550046078397705) (5000, 0.4528105007553442) (6000, 0.5162164387751635) (7000, 0.5683735945108044) (8000, 0.5569345984672456) (10000, 0.5959743436143418) (11000, 0.641709959717851) (12000, 0.6407842236996294) (13000, 0.6854715520191139) (15000, 0.7965583372646533) (16000, 0.7863989823017892)};
    \addplot[pattern=north west lines, draw=uchu-gray-6, pattern color=uchu-gray-6] fill between[of = Other and IPC];

    \coordinate (top) at (8000,1);
    \coordinate (hard) at (16000, 0.8);

    \end{axis}

    \node[annot x, minimum height=0, text height=0, anchor=south east] (annot) at ($(hard) + (2pt, 0)$) {\scriptsize +206\%};
    \node[above, text depth=0] at ($(top) + (0, 2pt)$) {\small Envoy};

\end{tikzpicture}
    \end{subfigure}%
    \caption{The eBPF runtime is less efficient than user space, such that the performance degrades more quickly when enforcing the Mutate policy. Despite this, \proj still outperforms the L4 fast path by $2.2\times$ in the worst case.}
    \label{fig:eval:pc:dissect}
\end{figure}
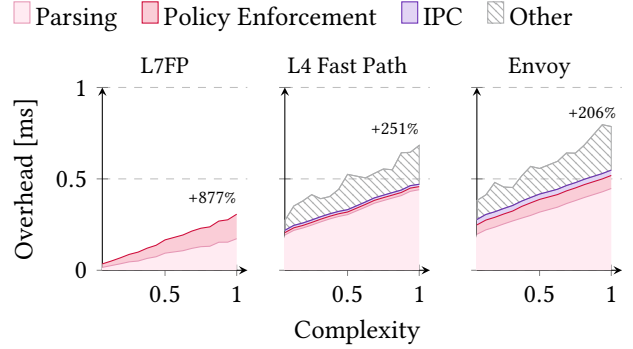

We illustrate this by measuring the main sources of overhead: parsing the message, enforcing the policy, and IPC. Note that we only measure the \emph{overhead} induced by the service proxy, and ignore the runtime of the application itself. It follows that \proj does not exhibit any IPC overhead, as it runs in kernel space. For each data point, the load generator sends 5K req/s for three minutes\footnote{We choose this rate because, in our setup, it exercises the full application without exceeding the maximum throughput that all configurations can process (see \Cref{fig:eval:pc:ssm}).} to collect enough samples. The result is shown in \Cref{fig:eval:pc:dissect}: While Envoy's overhead grows by 206\% (251\% with the L4 fast path), \proj grows by 877\%. It also becomes apparent that for Envoy, only the parsing overhead grows, whereas for \proj, the policy enforcement overhead grows at the same rate as parsing. Despite this, even for the maximum complexity $c=1$, \proj's overhead is only 0.30\,ms, whereas Envoy's overhead reaches 0.78\,ms for every request (0.68\,ms with the L4 fast path).

To summarize, eBPF's runtime is indeed less efficient than that of the user space. Therefore, offloading significantly more complex policies will inevitably lead to a point where the policy enforcement in eBPF is disadvantageous. However, we consider such a policy to be an extreme case that is unlikely to be used in practice.

\section{Discussion and Limitations}
\label{sec:discussion}

We start by discussing the future work that this project has made possible. Then we discuss the main limitations of \proj, and outline what it takes to alleviate them.

\paragraph{Accelerating other applications} This paper proposes a practical way to parse and process L7 protocols in the kernel. As long as the message processing is not too complex (c.f.~\Cref{sec:eval:cp}), \proj can significantly improve throughput and request latency. \proj's core ideas also apply to other applications that are configured in a similar way as service proxies. For example, application-level firewalls could profit heavily from an eBPF offload. Moreover, \proj could extend web applications to serve simple HTTP requests directly from kernel space, or cache more complex ones.


\paragraph{Parsing more complex protocols} The current version of \proj supports \hone and \htwo. Future versions can extend this support, e.g. for gRPC~\cite{online_grpc}, by constructing DFAs that operate on the respective encoding. In this regard, \hthree presents a special case. It uses QUIC, which the latest Linux kernel does not support. This renders it incompatible with the eBPF runtime. However, in-kernel support for QUIC is on the horizon~\cite{online_quic_kernel}.

\paragraph{Implementing more policies} \proj currently supports a handful of policies that are, according to our studies, commonly used in today's service meshes. Adding support for new policies is as simple as implementing a template. Our study has shown that 11\% of L7 policies are not fully compatible with the eBPF runtime and require kernel changes. Implementing one of such policies, e.g., a custom policy in the form of a Lua script, requires the user to additionally implement a kernel module that provides the Lua runtime to \proj's data plane. Standard eBPF techniques can be used to call the kernel module from within the template.

\paragraph{Enforcing more complex policies} \proj's data plane complexity is bound by the eBPF verifier. It limits the number of instructions an eBPF program may have, and fails to verify the safety of programs with too complex control flows. To address this, the eBPF community is actively working on increasing these limits~\cite{online_taking_bpf}, and academia has produced numerous works~\cite{vishwanathan_verifying, sun_validating, bhat_formal} that aim to improve the verifier's accuracy. \proj can also employ standard techniques like \emph{kfuncs} and tail calls to circumvent this limitation. We emphasize however, that for all experiments in this paper, \proj synthesizes a data plane that fits into a single eBPF program.

\section{Related Work}

\paragraph{Optimizing L4 policies} Cilium~\cite{online_cilium} and Calico~\cite{online_calico} are the de-facto industry standard for service meshes. They enforce L4 policies in kernel space, but remain dependent on user space service proxies to enforce L7 policies. We replicate their data path with the ``L4 Fast Path'' in \Cref{sec:eval:fp} and find that \proj achieves up to $3\times$ higher throughput. Other approaches reduce IPC overheads by bypassing the network stack~\cite{online_dpdk, rizzo_transparent, høiland-jørgensen_the, marinos_network, belay_ix}, or by batching the packets to be processed~\cite{online_io_uring}. Such approaches require application modifications -- \proj remains transparent. OnCache~\cite{lin_oncache} accelerates container overlay networks by caching tunnel headers in eBPF. Slim~\cite{zhuo_slim} is a container overlay network that implements lightweight network virtualization by manipulating connection-level metadata when connections are established. Spright~\cite{qi_spright} is a serverless framework that redirects traffic at the socket level. It uses shared memory to implement zero-copy message delivery. Deploying one sidecar per pod incurs a large overhead in the data plane due to the increased CPU utilization, but also makes the control plane more complex. Ambient~\cite{online_istioambient} addresses this by deploying a per-node L4 service proxy. \proj takes this one step further and removes, in many cases, the service proxy from the critical path. Finally, multiple works~\cite{moon_acceltcp, brunella_hxdp, shashidhara_flextoe} offload L4 policies like packet forwarding and filtering to programmable NICs. This requires specialized hardware whereas \proj improves performance with off-the-shelf hardware.

\paragraph{Optimizing L7 policies} Multiple works have advocated for application-layer in the kernel~\cite{sidoretti_application, shahinfar_automatic} or SDN architectures~\cite{antichi_fullstack}. \cite{kumar_feasibility}~studies the potential of eBPF to read L7 payloads. They find that eBPF-based telemetry is more efficient than state-of-the-art systems, but is also limited to matching only 48\,B of data. Nevertheless, processing L7 traffic in the kernel has remained impractical. Systems like strparser~\cite{online_strparser} facilitate message delineation within eBPF, and the kernel connection multiplexor (KCM)~\cite{online_kcm} leverage this to provide a message-based interface over TCP. Crucially, however, they are insufficient to parse L7 protocols with a self-describing structure, e.g. \http. \proj is based on strparser and able to match up to 16\,kB of data. \proj's parsing approach augments KCM to provide an HTTP-based message interface. \cite{ngai_regex} is another orthogonal approach to filter packets using regexes in eBPF. \proj employs a similar approach (Aho-Corasick DFAs) to parse messages efficiently. Copper~\cite{saxena_copper} proposes a policy DSL that facilitates the management of diverse data planes and helps minimize the data plane resources. CanalMesh~\cite{song_canal} implements a multi-tenant remote mesh gateway, eliminating the need for a sidecar proxy. This approach improves throughput and latency while simplifying orchestration and pod intrusion. Nevertheless, the deployment of a secure and error-free multi-tenant proxy can be impractical. Finally, mRPC~\cite{chen_remote} and ServiceRouter~\cite{saokar_servicerouter} eliminate the service proxy completely by reintegrating policy enforcement back into the application. This yields significant performance gains, but couples the life cycle of the pod to that of the policy enforcement.

\paragraph{Specializing the kernel.} There is a long line of work that aims at specializing the kernel~\cite{chick_shadow, pu_optimistic, Perianayagam_profile, pu_microlanguages}, or specifically the network stack~\cite{chen_a}. \cite{bhatia_automatic} propose an automatic specialization of protocol stacks, tailored towards the current usage context. \cite{marinos_network} employ Netmap~\cite{rizzo_transparent} to implement a specialized zero-copy network stack operating from user space. LinuxFP~\cite{abranches_linuxfp} continuously profiles the kernel to automatically synthesize and deploy a message processing fast path.


\section{Conclusion}

The service mesh simplifies the development of web applications by abstracting away the network layer. But its data plane, the service proxy, is responsible for major performance degradations, resulting in higher request latencies.

We presented \proj, a transparent kernel space fast path for service proxies. We demonstrated the need for application-layer processing in kernel space and showed that this has become a possibility with the latest advances in the Linux kernel. \proj improves the performance of service meshes dramatically, lowering the request latency while sustaining more throughput. The service proxy remains agnostic of \proj, allowing for a flexible deployment.

We consider \proj to be the first step that shows the performance benefits of application-layer processing in the kernel. This opens the door for performance optimization of many different applications that profit from application-level processing in the kernel.

\bibliographystyle{IEEEtran}
\bibliography{bib/library, bib/proj}

\end{document}